\documentclass[hyper]{JHEP3}
\usepackage{graphicx}

\input{epsf}
\usepackage{epsfig}
\usepackage{amssymb}
\usepackage{amsfonts}
\usepackage{amsbsy}
\usepackage{amsmath}
\usepackage{float}
\allowdisplaybreaks[2]

\usepackage{amssymb,amscd}
\usepackage{mathrsfs}
\usepackage{amsmath,amsthm}
\newtheorem*{conjecture}{Conjecture}

\newcommand{\CC}{\mathbb{C}} % Complessi
\newcommand{\RR}{\mathbb{R}} % Reali
\newcommand{\QQ}{\mathbb{Q}} % Razionali
\newcommand{\ZZ}{\mathbb{Z}} % Interi
\DeclareMathOperator{\Tr}{Tr} % traccia
\DeclareMathOperator{\SL}{SL} % special linear group
\DeclareMathOperator{\ch}{ch}

\newcommand{\N}{\mathcal{N}} % Andreotti-Mayer locus
\newcommand{\HH}{\mathbb{H}} % Upper half plane

\newcommand{\be}{\begin{equation}}
\newcommand{\ee}{\end{equation}}

\newcommand{\MM}{\mathbb{M}} % Mathieu group

\def\half{\frac{1}{2}}

\title{Mathieu Moonshine in the elliptic genus of K3}

\author{Matthias R.~Gaberdiel, Stefan Hohenegger and
Roberto Volpato
\\ ~
\\
Institut f\"ur Theoretische Physik \\
ETH Zurich \\
CH-8093 Z\"urich \\
Switzerland }

\abstract{It has recently been conjectured that the elliptic genus of K3 can be written in terms of
dimensions of Mathieu group $\MM_{24}$ representations. Some further evidence for this idea
was subsequently  found by studying the twining genera that are obtained from the elliptic genus
upon replacing dimensions of Mathieu group representations by their characters.
In this paper we find explicit formulae for all (remaining) twining genera by making
an educated guess for their general modular properties. This allows us to identify the decomposition of
{\em all} expansion coefficients in terms of dimensions of $\MM_{24}$-representations.
For the first 500 coefficients we verify that the multiplicities with which these representations appear
are indeed all non-negative integers. This represents very compelling evidence in favour of the
conjecture.}

\begin{document}

\section{Introduction}

A few months ago Eguchi, Ooguri and Tachikawa \cite{EOT} observed that the elliptic genus
of K3 seems to involve representations of the largest of the Mathieu groups,
$\MM_{24}$. More specifically, they studied the expansion of the elliptic genus of K3 in terms of
elliptic genera of $\mathcal{N}=4$ superconformal representations following \cite{EH}, and
noted that the expansion coefficients can be written in terms of dimensions of
representations of $\MM_{24}$. This intriguing observation is very reminiscent of a similar
phenomenon usually referred to as `Monstrous Moonshine', namely  that the famous $J$-function
has an expansion in terms of characters of the Virasoro algebra whose coefficients are
dimensions of Monster group representations, as was first noted by McKay and Thompson.

In the context of Monstrous Moonshine, this observation was eventually explained by the
construction of the so-called Monster conformal field theory $V^\natural$ \cite{FLM}, a self-dual
conformal field theory at $c=24$ whose space of states is of the form
\begin{equation}\label{MCFT}
V^\natural = \bigoplus_{n=0}^{\infty} \left(V_n \otimes {\cal H}^{\rm Vir}_{h=n} \right) \ ,
\end{equation}
where each $V_n$ is a representation of the Monster group, while ${\cal H}^{\rm Vir}_{h}$
denotes the irreducible Virasoro representation with conformal weight $h$ and $c=24$.
The $J$-function is then the partition function of $V^\natural$, and its Fourier coefficients are
therefore sums of dimensions of irreducible Monster group representations. Furthermore,
the automorphism group of $V^\natural$ is the Monster group.

One key observation that provided convincing evidence for  (\ref{MCFT}) came from
considering the so-called McKay-Thompson series \cite{Thompson}. These are obtained from
the $J$-function upon replacing the expansion coefficients $A_n=\dim(V_n)$  by their corresponding
characters, $\Tr_{V_n}(g)$, where $g$ is an element of the Monster group.\footnote{These series can be
calculated provided one knows how to write the Fourier coefficients of the $J$-function
as $A_n=\dim(V_n)$, but do not require any additional knowledge.} It was
shown by Conway \& Norton \cite{CN} that these McKay-Thompson series have nice modular properties
under congruence  subgroups of $SL(2,\mathbb{Z})$. This is what one expects if they
arise indeed from (\ref{MCFT}) since they are then equal to the `twining character', {\it i.e.}\ the
character with the insertion of the group element $g$, which has good modular properties based
on standard orbifold arguments. For a review of these and other aspects of Monstrous Moonshine see
{\it e.g.}~\cite{Gannon}.
\smallskip

By analogy to (\ref{MCFT}) the observation of \cite{EOT} suggests
that the states that contribute to the elliptic genus
of K3 have the structure
\begin{equation}\label{EOT}
{\cal H}^{\rm BPS} = \bigoplus_{n} \left( H_n \otimes {\cal H}^{{\cal N}=4}_{n} \right) \ ,
\end{equation}
where the sum runs over all irreducible ${\cal N}=4$ representations that contribute to the
elliptic genus, while each $H_n$ is a $\MM_{24}$ representation. In order to test this
idea it is then again natural to consider the `twining genera' $\phi_g(\tau,z)$, where one replaces
$\dim(H_n)$ by the corresponding character $\Tr_{H_n}(g)$, with $g\in \MM_{24}$
\cite{Cheng,Gaberdiel:2010ch}.
Unfortunately,  the dimensions of the irreducible representations of $\MM_{24}$ are rather small, and
the decomposition of the expansion coefficients of the elliptic genus of K3 in terms of
dimensions of $\MM_{24}$ representations could only be guessed reliably for the first few coefficients.
However, the elliptic genus of K3 is a weak Jacobi form of index one and weight zero
\cite{Kawai:1993jk}, and one expects that the twining genera  should have similar properties.
More specifically, if $g\in \MM_{24}$ has order $N$, then the corresponding twining genus
$\phi_g(\tau,z)$ should transform as a weak Jacobi form of index $1$ and weight $0$ under the
congruence subgroup $\Gamma_0(N)$, possibly up to a multiplier system.

In \cite{Cheng,Gaberdiel:2010ch} this assumption about the modular properties
of the twining genera, together with the explicit knowledge of the first few coefficients, was used
in order to determine some of them explicitly. The fact that these two
constraints were compatible was already a fairly non-trivial consistency check on the proposal
of \cite{EOT}. Furthermore, for group elements that enjoy an interpretation as an automorphism
symmetry of K3 at a suitable point in moduli space, one could calculate the corresponding twining
genera directly (see in particular \cite{David:2006ji}). In this way some of these formulae could be
confirmed independently \cite{Cheng}.
\smallskip

In this paper we shall complete the analysis of \cite{Cheng,Gaberdiel:2010ch} by finding explicit
formulae for the remaining conjugacy classes.\footnote{After posting the first version of this paper
on the arXiv, we were informed by Tohru Eguchi that they have independently obtained explicit formulae
for these twining genera which agree with ours, see \cite{EHT}.}
In particular, we shall make a precise proposal
for the structure of the multiplier system that appears in the modular transformation formula, see
(\ref{conj}) and table~\ref{t:hvalues} below. We can then follow again the strategies outlined in
\cite{Gaberdiel:2010ch} and \cite{Cheng}, respectively, to determine the twining genera, using the
explicit knowledge of the first few terms. The fact that this procedure is successful is again a fairly
non-trivial consistency check of the proposed structure (\ref{EOT}).

The knowledge of {\em all} twining genera also leads to another, highly constraining, consistency
check. If \eqref{EOT} holds and if, for each $g$, our explicit formula for $\phi_g$ really corresponds
to the trace over ${\cal H}^{\rm BPS}$ with the insertion of $g$, then its coefficients must be equal to
$\Tr_{H_n}(g)$ for all $n$. But knowing these traces for all classes $g$ is sufficient to identify
unambiguously $H_n$ as a (possibly reducible) representation of $\MM_{24}$. Thus assuming that
(\ref{EOT}) holds,  we can {\em deduce} the decomposition of  $H_n$ into irreducible $\MM_{24}$
representations for all $n$.
We have calculated the multiplicities explicitly up to order $500$ --- the results up to order $30$ are
tabulated in table~\ref{t:decom} --- and they are all indeed non-negative integers. The fact that this
decomposition works out, {\it i.e.}\ that each $H_n$ can be written as a direct sum
of irreducible representations (with integer multiplicities) is then a highly non-trivial check of
(\ref{EOT}); in fact, we would in some sense prove the proposal of \cite{EOT} if we could show that
{\em all} of these multiplicities are indeed non-negative integers.
\medskip

The paper is organised as follows. In section~\ref{s:elliptic} we briefly review some of the properties of the
elliptic genus of K3, as well as the proposal of \cite{EOT}. The modular properties of the
twining genera are explained in detail in section~\ref{s:modprop}. In particular, by analogy with the
situation for the McKay-Thompson series of Monstrous Moonshine, we make a specific proposal for the
structure of the multiplier system. This proposal is then subsequently verified in sections~3.1 and
3.2. In section~3.1 we adopt the strategy of \cite{Gaberdiel:2010ch} and consider the twining genus
at specific values of $z$, finding explicit formulae for all remaining cases.
Section~3.2, on the other hand,
employs the method of \cite{Cheng} to write the twining genus in terms of a modular form of weight two;
this allows us to check some of the modular properties of the twining genera fairly directly, but the
analysis is quite complicated and the details are described in appendix~B and C. Finally, we
explain in section~\ref{s:tables} how the knowledge of all twining genera determines the decomposition
of the coefficients in terms of $\MM_{24}$ representations, and give the explicit results in table~\ref{t:decom}.
We close with some comments and speculations in section~\ref{Sect:Conclusions}.

%%%%%%%%%%%%%%%%%%%%%%%%%%%%%%%%%%%%%%%%%%%%%%%%%%%%%%%%%%%%%%%%%%%
\section{Elliptic genus and twining characters}\label{s:elliptic}

The elliptic genus of an ${\cal N}=2$ superconformal algebra is defined by
\begin{equation}
\phi(\tau,z) = \hbox{Tr}_{{\cal H}_{\text{RR}}} \Bigl( q^{L_0-\frac{c}{24}} e^{2\pi i z J_0} (-1)^F \,
\bar{q}^{\bar{L}_0-\frac{\bar{c}}{24}} (-1)^{\bar{F}} \Bigr) \ ,
\end{equation}
where $q=e^{2\pi i\tau}$, and the trace is taken in the RR sector. For the right-movers
(whose modes are denoted by a bar), only the ground states contribute,
and hence the above expression is in fact independent of $\bar{q}$.
As is well known \cite{Kawai:1993jk}, the  modularity properties of
conformal field theory together with spectral flow invariance and unitarity imply
that the elliptic genus is a \emph{weak Jacobi form} of index $m=\frac{c}{6}$
and weight $0$ \cite{EichlerZagier}.
A weak Jacobi form $\phi(\tau,z)$ of weight $w$ and index $m\in{\mathbb Z}$
is a function $\phi$ of $(\tau,z)\in \HH\times \CC$, where $\HH$ is the upper half-plane.
It is characterised by the transformation properties
\begin{equation}\label{eq:jactmn1}
 \phi\Bigl(\frac{a \tau + b}{c \tau + d} , \frac{z}{c \tau + d}\Bigr) =
(c \tau+d)^w \, e^{ 2 \pi i m \frac{c z^2}{c \tau + d} } \, \phi(\tau,z)
\qquad \begin{pmatrix} a & b \\ c & d \\ \end{pmatrix} \in SL(2,\ZZ) \ ,
\end{equation}
\begin{equation}\label{eq:jactmn2}
 \phi(\tau,z+ \ell \tau + \ell') = e^{-2 \pi i m
(\ell^2 \tau+ 2 \ell z)} \phi(\tau,z) \qquad\qquad\qquad
\ell,\ell'\in \ZZ \ ,
\end{equation}
and has a Fourier expansion
\begin{equation}
 \phi(\tau,z) = \sum_{n \geq 0, \ell\in \ZZ} c(n,\ell) q^n
y^\ell \end{equation}
where $y= e^{2\pi i z}$ and $c(n,\ell) = (-1)^w c(n,-\ell)$.
For the case of K3 that will concern us primarily in this paper, $m=1$ and the
elliptic genus equals \cite{EOTY}
\begin{equation}\label{K3R}
\phi_{K3}(\tau,z)   =
2 y + 20 + 2 y^{-1} + q \Bigl(20 y^2 -128 y + 216 - 128 y^{-1} + 20 y^{-2} \Bigr) + {\cal O}(q^2) \ .
\end{equation}
It can be thought of as the partition function of the ${\cal N}=2$  half-BPS states of type II
string theory on K3.

For the case of K3 the conformal field theory is actually ${\cal N}=4$ superconformal, and
one can therefore write the elliptic genus in terms of the elliptic genera associated to
${\cal N}=4$ superconformal representations. It was observed in \cite{EOT}, following on
from earlier work \cite{EH}, that it can be written as
\begin{equation}\label{3.1}
\phi_{K3}(\tau,z) = 24 \ch_{h=\frac{1}{4},l=0}^{\, {\cal N}=4} (\tau,z)
+ \sum_{n=0}^{\infty}  A_n \ch_{h=n+\frac{1}{4},l=\frac{1}{2}}^{\, {\cal N}=4} (\tau,z)\ ,
\end{equation}
where  $\ch_{h=\frac{1}{4},l=0}^{\, {\cal N}=4}$ is the elliptic genus of the short ${\cal N}=4$
representation with $h=\frac{1}{4}$ and $l=0$ --- see \cite{Eguchi:1987wf,Eguchi:1988af} for
an explicit formula ---  while
\begin{equation}\label{long}
\ch_{h,l=\frac{1}{2}}^{\, {\cal N}=4} (\tau,z) = q^{h-\frac{3}{8}}\,  \frac{\vartheta_1(\tau,z)^2}{\eta(\tau)^3}
\end{equation}
is the elliptic genus of a long ${\cal N}=4$ representation.\footnote{Strictly speaking, the ${\cal N}=4$
representation with $n=0$ ($h=\tfrac{1}{4}$) is short, and thus (\ref{long}) for $h=\tfrac{1}{4}$ is not
the elliptic genus of a single representation, but rather involves a sum of representations.} The observation
of \cite{EOT} was that the coefficients $A_n$ can be written in terms of dimensions of
representations $H_n$ of the Mathieu group ${\mathbb M}_{24}$, so that
\begin{align}
\phi_{K3}(\tau,z)=&(\dim H_{00})\ch^{\N=4}_{h=\frac{1}{4},l=0}(\tau,z)
-(\dim H_0)\ch_{h=\frac{1}{4},l=\frac{1}{2}}^{\, {\cal N}=4} (\tau,z)\notag\\&+\sum_{n=1}^\infty(\dim H_n)\ch_{h=n+\frac{1}{4},l=\frac{1}{2}}^{\, {\cal N}=4} (\tau,z)\ ,\label{ellrep}
\end{align}
where
\begin{equation}\label{3.3}\begin{array}{rclrcl}
H_{00} & = & {\bf 23} + {\bf 1} \qquad \qquad
&  H_0 & = &  2\cdot {\bf 1}  \\
H_1 & = & {\bf 45} +\overline{\bf 45}  \qquad \qquad
& H_2 & = & {\bf 231} + \overline{\bf 231} \\
H_3 & = & {\bf 770} + \overline{\bf 770} \qquad \qquad
& H_4 & = & {\bf 2277} + \overline{\bf 2277} \\
H_5  & = & 2\cdot {\bf 5796}  \qquad \qquad
& H_6& = & 2\cdot {\bf 3520} + 2\cdot {\bf 10395}\\
\end{array}
\end{equation}
\be\label{3.4}
H_7  =  2\cdot {\bf 1771} + 2 \cdot {\bf 2024} + 2\cdot {\bf 5313} + 2\cdot {\bf 5796}
+ 2\cdot {\bf 5544} + 2\cdot {\bf 10395}\ .
\ee
The dimensions of the irreducible representations of ${\mathbb M}_{24}$ can be read off
from the character table (see table~\ref{t:char}). Note that we have absorbed the prefactor
$2$ in equation (1.11) of \cite{EOT} into the definition of $A_n=\dim(H_n)$. Then we can write
the $H_n$ in terms of real representations, so that, for example, $H_1$ is the sum of a pair of conjugate representations. The expression for $H_7$ was given in \cite{Cheng,Gaberdiel:2010ch}
and differs from what was originally proposed in \cite{EOT}.

It is natural to conjecture that such a decomposition is the hallmark of a deeper structure underlying
the elliptic genus of K3, see (\ref{EOT}). In order to test this idea, the `{\em twining elliptic genera}',  {\it i.e.}\ the
analogues of the McKay-Thompson series  of Monstrous Moonshine, were
considered in \cite{Cheng,Gaberdiel:2010ch}. These twining genera are obtained from
the elliptic genus upon inserting a group element $g\in {\mathbb M}_{24}$ into the
trace
\begin{equation}
\phi_g(\tau,z) = \frac{1}{2}\Tr_{{\cal H}_{\text{RR}}}\Bigl( g\, q^{L_0-\frac{c}{24}} e^{2\pi i z J_0} (-1)^F \,
\bar{q}^{\bar{L}_0-\frac{\bar{c}}{24}} (-1)^{\bar{F}} \Bigr) \ .
\end{equation}
As in \cite{Gaberdiel:2010ch}, we shall normalise the twining characters so that
$\phi_{\rm 1A}(\tau,z) = \frac{1}{2} \phi_{K3}(\tau,z)$ is directly equal to the standard weak
Jacobi form $\phi_{0,1}$ (see appendix \ref{App:Definitions}).
Technically speaking, $\phi_g$ is simply obtained from \eqref{ellrep} by replacing the
dimensions $A_n=\dim(H_n)$ by the trace of $g$ over $H_n$,
$A_n(g)=\Tr_{H_n}(g)$, {\it i.e.}\
\begin{align}
\phi_{g}(\tau,z)=\frac{1}{2}\Bigl[\Tr_{H_{00}}&(g)\ch^{\N=4}_{h=\frac{1}{4},l=0}(\tau,z)
-\Tr_{H_0}(g)\ch_{h=\frac{1}{4},l=\frac{1}{2}}^{\, {\cal N}=4} (\tau,z)
\nonumber\\
&+\sum_{n=1}^\infty\Tr_{H_n}(g)\ch_{h=n+\frac{1}{4},l=\frac{1}{2}}^{\, {\cal N}=4} (\tau,z)\Bigr]\ .\label{phigH}
\end{align}
The character of $g$ only depends on its conjugacy class, and thus the various traces can be
read off from the character table of  ${\mathbb M}_{24}$, see table~\ref{t:char}.

As discussed  in more detail in the next section, the twining genera are expected to be
Jacobi forms under suitable congruence subgroups of $\SL(2,\ZZ)$. This was confirmed
for a number of conjugacy classes in \cite{Cheng,Gaberdiel:2010ch}.
In the next section we shall complete this programme by determining the twining characters
for all remaining conjugacy classes. We shall furthermore show that they have the appropriate
modular properties.

\section{Modular properties of the twining genera}\label{s:modprop}

Using standard conformal field theory arguments, it was argued in \cite{Cheng,Gaberdiel:2010ch}
that the twining genera $\phi_g$ should transform as  Jacobi forms of index $1$ and weight $0$,
possibly up to a phase, under the congruence subgroup $\Gamma_0(N)$. More specifically, this means
that $\phi_g$ satisfies (\ref{eq:jactmn2}) for all $\ell,\ell'\in {\mathbb Z}$, while
(\ref{eq:jactmn1}) only holds for
\be
\begin{pmatrix}a & b\\ c & d\end{pmatrix} \in
\Gamma_0(N)=\left\{\begin{pmatrix}a & b\\ c & d\end{pmatrix}\in \SL(2,\ZZ)\mid c\equiv 0\mod N\right\}\ ,
\ee
where $N$ is the order of $g$. In \cite{Cheng}, this was explicitly verified for the conjugacy
classes\footnote{The classes which are power conjugated, for example 7A and 7B,
give rise to the same twining genus, so that we denote them as a unique class 7AB. The
twining genera for the classes 1A, 2A, 3A, 4B, 5A and 6A were also found in \cite{Gaberdiel:2010ch}.}
\be
\label{set1} {\rm 1A},\ {\rm 2A},\ {\rm 3A},\ {\rm 4B},\ {\rm 5A},\
{\rm 6A},\ {\rm 7AB},\ {\rm 8A},\ {\rm 11A},\ {\rm 14AB},\ {\rm 15AB},\ {\rm 23AB}\ .
\ee
These classes are characterised by the condition $\phi_g(\tau,0)\neq 0$. This is equivalent to the condition
that  a representative of the class is contained in the subgroup $\MM_{23}\subset \MM_{24}$,
where we think of $\MM_{24}$ as a subgroup of $S_{24}$, the permutation group of $24$ points,
and define $\MM_{23}\subset \MM_{24}$ to be the subgroup fixing, say, the first point.
All geometric symmetries of K3 at a suitable point in moduli space lie in this subgroup
\cite{Mukai,Kondo}, and thus some of them can be calculated from first principles, see also \cite{David:2006ji}.
For all of them the multiplier system turned out to be trivial \cite{Cheng}.
The situation is more difficult for the remaining conjugacy classes
\be
\label{set2} {\rm 2B},\ {\rm 3B},\ {\rm 4A},\ {\rm 4C},\ {\rm 6B},\ {\rm 10A},\ {\rm 12A},\
{\rm 12B}, \ {\rm 21AB}\ ,
\ee
since there is no {\it a priori} method to determine them. In \cite{Gaberdiel:2010ch},
explicit formulae were found for the first few of them by combining the constraints from modularity with
the knowledge of the first few coefficients.\footnote{The twining characters for 2B and 4A were also found
in \cite{Cheng}.} In fact, the analysis of  \cite{Gaberdiel:2010ch} was performed for the NS-sector
version of the twining genus, the twining character
\begin{equation}
\chi_g(\tau,z) = \exp\left[2 \pi i  \left(\frac{\tau}{4} + z+ \half \right) \right]
\phi_g\left(\tau, z + \frac{\tau}{2} + \half  \right)
\end{equation}
evaluated at $z=0$. The advantage of this approach is that one can work with standard
modular functions (rather than Jacobi forms). The price one has to pay, on the other hand,
is that part of the modular invariance is broken, and that multiplier phases are introduced
for certain modular transformations. The latter property turned out to be a blessing in disguise
since it suggested that multiplier phases may naturally appear in the modular transformations. Indeed,
in \cite{Gaberdiel:2010ch} the twining characters were determined for all elements $g$ up to order
$o(g)\le 6$, and it was found that the classes  in \eqref{set1} and in \eqref{set2} appear to behave
very similarly. This suggests that also the twining genera $\phi_g$ associated to  \eqref{set2}
should be invariant under $\Gamma_0(N)$, possibly up to non-trivial phases.

The appearance of a multiplier system in the transformation rule for the twining genera is certainly
consistent with standard CFT arguments. In fact, this phenomenon also occurs for several
McKay-Thompson series (and, more generally, for replicable functions \cite{FMN}), which are the
analogues of the twining characters in the context of Monstrous Moonshine \cite{CN}. Recall that
each McKay-Thompson series $T_g$ is associated with a certain discrete group
\be
\Gamma_0(N|h)=\left\{\begin{pmatrix}a & b/h\\ Nc & d\end{pmatrix}\in \SL(2,\RR)\mid a,b,c,d\in\ZZ\right\}\ ,
\ee
where $N$ is the order of the Monster class $g$, and $h$ is some integer such that
$h|\gcd(N,24)$. This group (sometimes with the inclusion of some Atkin-Lehner involutions of
$\Gamma_0(N)$, see \cite{CN}) is the (restricted) eigengroup of $T_g$, {\it i.e.}\ the group under
which $T_g$ is invariant up to $h$-th roots of unity. In particular, $T_g$ is invariant (without any phases)
under $\Gamma_0(Nh)\subset \Gamma_0(N)$, while under the cosets of
$\Gamma_0(N|h)/\Gamma_0(Nh)$, that are represented by
\begin{equation}\label{cosets}
\left(\begin{matrix}1& 1/h\\ 0 & 1\end{matrix}\right) \qquad \text{and}\qquad
\left(\begin{matrix}1& 0\\ N & 1\end{matrix}\right) \ ,
\end{equation}
it transforms as
\begin{align}\label{phase1}
T_g\Bigl(\tau+\frac{1}{h}\Bigr)=e^{-\frac{2\pi i}{h}}T_g(\tau)\qquad \text{and} \qquad
T_g\Bigl(\frac{\tau}{N\tau+1}\Bigr)=e^{\pm\frac{2\pi i}{h}}T_g(\tau)\ .
\end{align}
The two cosets in (\ref{cosets})  generate $\Gamma_0(N|h)$, so that
\eqref{phase1} uniquely determines the multiplier system under $\Gamma_0(N|h)$.

It is then natural to expect that  analogous properties hold for the twining genera of $\MM_{24}$.
The most obvious generalisation would be to require the twining genus $\phi_g$, with $g\in \MM_{24}$
and $o(g)=N$, to be a Jacobi form (with a suitable multiplier system) of weight $0$ and index
$1$ under $\Gamma_0(N|h)$, for some $h|\gcd (N,24)$. However, there is one immediate problem
with this proposal:
for $h>1$, $\Gamma_0(N|h)$ is not contained in $\SL(2,\ZZ)$, and it is not clear how to
define the action of the whole $\Gamma_0(N|h)$ on Jacobi forms.\footnote{A well-defined action of
$\SL(2,\RR)$ on Jacobi forms can be defined, see \cite{EichlerZagier}. However, this action does
not respect the periodicity condition on $z$, and thus does not seem to be relevant in the
current context.} Thus we can only analyse the modular properties under the subgroup
$\Gamma_0(N|h)\cap \SL(2,\ZZ)\cong \Gamma_0(N)$. This then leads to the following conjecture:

\begin{conjecture}
For all the conjugacy classes $g$ of $\MM_{24}$, the twining character $\phi_g(\tau,z)$ is a
Jacobi form of index one and weight zero under $\Gamma_0(N)$, with a multiplier system
defined by\footnote{We thank Miranda Cheng and John Duncan for pointing out an error in a previous version of this formula.}
\be\label{conj} \phi_g\Bigl(\frac{a\tau+b}{c\tau+d},\frac{z}{c\tau+d}\Bigr)
=e^{\frac{2\pi i cd}{Nh}}\,e^{\frac{2\pi i\, cz^2}{c\tau+d}}\,\phi_g(\tau,z)\ ,\qquad
\begin{pmatrix}a & b\\ c&d \end{pmatrix}\in \Gamma_0(N)\ ,
\ee
where $N$ is the order of $g$ and $h|\gcd(N,12)$. The multiplier system is
trivial ($h=1$) if and only if $g$ contains a representative in $\MM_{23}\subset \MM_{24}$.
\end{conjecture}

For the classes in \eqref{set1} that have representatives in $\MM_{23}$ the conjecture has been
shown in \cite{Cheng}. In the next subsections, we will show that the conjecture is also true for
the remaining classes, {\it i.e.}\ the elements in \eqref{set2}, with the values of $h$ as given
in table \ref{t:hvalues}.  \begin{table}[H]\centerline{
$\begin{array}{|c|ccccccccc|}\hline
\text{Class}  & {\rm 2B} & {\rm 3B} & {\rm 4A} & {\rm 4C} & {\rm 6B}
& {\rm 10A} & {\rm 12A} & {\rm 12B} & {\rm 21AB} \\ \hline
h  & 2 & 3 & 2 & 4 & 6 & 2 & 2 & 12 & 3\\ \hline
\end{array}$}\caption{Value of $h$ for the conjugacy classes in (3.3).}\label{t:hvalues}
\end{table}
Because of these non-trivial multiplier systems, the analysis is quite difficult, and we have
applied two different strategies. First we have refined the method of
\cite{Gaberdiel:2010ch} by considering the twining genus $\phi_g(\tau,z)$ as a function of $\tau$ at
special  values of $z$. These values are chosen in such a way that the $z$-dependent exponential
factor  in the transformation formula \eqref{conj} cancels (part of) the $h$-th root of unity.  If the
phase is completely removed (as is the case for all but two classes), the resulting function is  a
modular function for $\Gamma_0(N)$, which can be easily analysed. Using this approach (as well
as some guess work for the other two cases)
we have  succeeded  in finding closed formulae for all the characters with $h>1$, see section~\ref{s:fixz}.

The other strategy follows the idea advocated in \cite{Cheng} and consists of expanding
the twining genus in terms of standard weak Jacobi forms. This reduces the problem to finding a
suitable modular form of weight two. This problem can be studied systematically, but usually leads to
more complicated computations. In section~\ref{s:wei2} we shall explain the salient features of this
analysis, while the explicit formulae for all characters are given in appendix~\ref{a:weightwo}.

\subsection{Characters evaluated at fixed values of $z$}\label{s:fixz}

Let $\phi(\tau,z)$ be a weak Jacobi form of weight $0$ and index $1$, transforming as in
\eqref{conj}, for some $N$ and $h$ with $h|\gcd(N,12)$. Then, for any $k\in\ZZ$, we have
\begin{align} \phi\Bigl(\frac{a\tau+b}{c\tau+d},\frac{k}{h}\Bigr)
=&\ \phi\Bigl(\frac{a\tau+b}{c\tau+d},\frac{k(c\tau+d)}{h(c\tau+d)}\Bigr)
=e^{\frac{2\pi i cd}{Nh}} \, e^{\frac{2\pi i\, k^2c(c\tau+d)}{h^2}}\, \phi\Bigl(\tau,\frac{k(c\tau+d)}{h}\Bigr)\notag\\
=& \ e^{\frac{2\pi i cd}{Nh}}\, e^{\frac{2\pi i  k^2c(c\tau+d)}{h^2}}
\, e^{-2\pi i (\frac{k^2c^2}{h^2}\tau+2\frac{k^2cd}{h^2})} \,
\phi\Bigl(\tau,\frac{kd}{h}\Bigr)\\
=& \ e^{2\pi i cd(-\frac{k^2 }{h^2} +\frac{1}{Nh})}\,
\phi\Bigl(\tau,\frac{kd}{h}\Bigr)\ ,\hspace{1.5cm}\text{for}\hspace{0.5cm} \begin{pmatrix}a & b\\ c& d\end{pmatrix}\in \Gamma_0(N)\ ,\notag
\end{align}
where we have used (\ref{eq:jactmn2})  in the second line. (Note that $c\in N{\mathbb Z}$, and hence
$\tfrac{c}{h}\in{\mathbb Z}$.)
Let us define
\be\label{defPhi} \Phi^{(h)}(\tau)=\frac{1}{\varphi(h)}\sum_{k\in (\ZZ/h\ZZ)^*}\phi\Bigl(\tau,\frac{k}{h}\Bigr)\ ,
\ee
where $(\ZZ/h\ZZ)^*$ is the set of totatives of $h$, {\it i.e.}\  the  positive integers smaller
than $h$ that are relatively prime to $h$, and the Euler totient function $\varphi(h)$ is the
number of all such totatives. This definition simplifies considerably in concrete
examples
\begin{align}
\Phi^{(h)}(\tau)=&\phi(\tau,\tfrac{1}{h})\ ,&& h=2,3,4,6\ ,\\
\Phi^{(12)}(\tau)=&\frac{1}{2}\bigl(\phi(\tau,\tfrac{1}{12})+\phi(\tau,\tfrac{5}{12})\bigr)\ ,&&
\end{align}
because $\phi(\tau,z)=\phi(\tau,-z)$ for Jacobi forms of even weight.
It is easy to verify that, for all $h|12$, the condition $\gcd (k,h)=1$ implies $k^2\equiv 1\mod h$.
Furthermore, for  $(\begin{smallmatrix}a & b\\ c& d\end{smallmatrix})\in \Gamma_0(N)$,
the condition $ad-bc=1$ implies $\gcd(d,h)=1$, so that the map $k\mapsto kd$ is
bijective on $(\ZZ/h\ZZ)^*$. Thus we conclude
\be\label{3.13}
 \Phi^{(h)}\Bigl(\frac{a\tau+b}{c\tau+d}\Bigr)=
e^{2\pi i cd(-\frac{ 1}{h^2}+\frac{1}{Nh})}\,
\Phi^{(h)}(\tau)\ ,\qquad\qquad \begin{pmatrix}a & b\\ c& d\end{pmatrix}\in \Gamma_0(N)\ ,
\ee
and it follows immediately that $\Phi^{(h)}(\tau)$ is invariant under $\Gamma_0(Nh)$.
Since $h$ divides $N$, the cosets of $\Gamma_0(N)/\Gamma_0(Nh)$ are generated by
$\left(\begin{smallmatrix}1&0\\N&1\end{smallmatrix}\right)$, and the phase in (\ref{3.13})
cancels under this transformation if and only if
\be\label{condi}
\frac{N}{h}\equiv 1\mod h\ .
\ee
Thus $\Phi^{(h)}$ is actually invariant under $\Gamma_0(N)$ if (\ref{condi}) holds. In this case, it
is easy to obtain a closed formula for $\Phi^{(h)}_g$. In fact, with the exception of $N=21$,
all the groups $\Gamma_0(N)$ we are interested in are genus zero\footnote{Genus zero
here means that the Riemann surface obtained by quotienting the upper half-plane
$\HH$ by $\Gamma_0(N)$ has the topology of the sphere.}, so that all modular functions
must be rational functions of the corresponding Hauptmodul.

As it turns out the condition (\ref{condi}) is satisfied for all classes in \eqref{set2}, with the exception
of 4A and 12A for which $h=2$ and hence $N/h \equiv 0$ mod $h$. In all other cases
we found a modular function for $\Gamma_0(N)$ which matches the first few coefficients in the
$q$-expansion that can be determined from (\ref{3.3}) and (\ref{3.4}).
This function is either a constant or a fractional linear transformation of the Hauptmodul
(in fact, a McKay-Thompson series) for $\Gamma_0(N)$. For the class 21AB it is a rational function
in the McKay-Thompson series $T_{\rm [21B]}$ and $T_{\rm [21D]}$, that are modular functions for
$\Gamma_0(21)$ (for a $q$-expansion of these McKay Thompson series see appendix~\ref{App:Definitions}).
Our explicit expressions are:
\begin{align}\Phi_{{\rm 2B}}^{(2)}(\tau)&=-4&& h=2\notag\\
\Phi_{{\rm 3B}}^{(3)}(\tau)&=-3 & &h=3\notag\\
\Phi_{{\rm 4C}}^{(4)}(\tau)&=-2 & &h=4\notag\\
\Phi_{{\rm 6B}}^{(6)}(\tau)&=-1 & &h=6\label{formPhi}\\
\Phi_{{\rm 10A}}^{(2)}(\tau)&=-4\frac{\eta(5\tau)\eta(2\tau)^5}{\eta(10\tau)\eta(\tau)^5}
=\frac{-20}{T_{\rm[10E]}-3}-4
& &h=2\notag\\
\Phi_{{\rm 12B}}^{(12)}(\tau)&=
-2-6\frac{\eta(2\tau)^2\eta(3\tau)\eta(12\tau)^3}{\eta(\tau)^3\eta(4\tau)\eta(6\tau)^2}
=\frac{-6}{T_{\rm[12I]}-3}-2 & &h=12 \notag\\
\Phi_{21AB}^{(3)}(\tau)&=\frac{1}{2}-\frac{7}{2}\frac{\eta(3\tau)\eta(7\tau)^3}{\eta(\tau)^3\eta(21\tau)}
= \frac{-T_{\rm [21B]}+7T_{\rm [21D]}+15}{-2T_{\rm [21B]}+2}& &h=3\ . \notag
\end{align}
Since $\phi_g(\tau,z)$ is a Jacobi form of index one, we have
\begin{equation}\label{rel2g}
 \phi_g(\tau,z)=\phi_g(\tau,0)\, \frac{\vartheta_2(\tau,z)^2}{\vartheta_2(\tau,0)^2}
+\phi_g(\tau,\tfrac{1}{2})\, \frac{\vartheta_1(\tau,z)^2}{\vartheta_2(\tau,0)^2} \ ,
\end{equation}
so that \eqref{formPhi} immediately gives a formula for the corresponding character at generic $z$
\be
\phi_g(\tau,z)=\Phi_g^{(h)}(\tau)\, \varphi(h)\,
\biggl( \sum_{k\in (\ZZ/h\ZZ)^*}\vartheta_1\Bigl(\tau,\frac{k}{h}\Bigr)^2 \biggr)^{-1}\, \vartheta_1(\tau,z)^2\ ,
\ee
where we also used the fact that, for all $g$ in \eqref{set2},
\be \phi_g(\tau,0)=\frac{1}{2}\Tr_{\mathbf{23}\oplus\mathbf{1}}(g)=0\ .
\ee
\smallskip

The remaining two cases, 4A and 12A, are not invariant under $\Gamma_0(N)$, and thus require
more work. For 4A a closed formula was already found in
\cite{Cheng,Gaberdiel:2010ch}
\be \phi_{\rm 4A}(\tau,\tfrac{1}{2}) =-4-32\frac{\eta(2\tau)^2\eta(8\tau)^4}{\eta(\tau)^4\eta(4\tau)^2}=-4-\frac{32}{T_{8E}(\tau)+4}
% -{\displaystyle \left(\frac{\eta(\tau) \,
%\eta(2\tau)}{\eta(\frac{\tau}{2}) \, \eta(4\tau)}\right)^4\frac{\vartheta_2(\tau)^2}
%{\vartheta_3(\tau)^2}}
\qquad h=2\ ,
\ee
and it is easy to verify that the corresponding $\phi_{\rm 4A}(\tau,z)$
transforms as in \eqref{conj} under $\Gamma_0(4)$ with $h=2$. We have also
managed to find a closed formula for the NS-character $\chi_g$  of 12A at $z=\tfrac{1}{6}$
\begin{align}
 \chi_{{\rm 12A}}(\tau,\tfrac{1}{6})=&\frac{\eta(\tau)\eta(\frac{3\tau}{2})^2\eta(4\tau)^2\eta(6\tau)^3}{\eta(2\tau)
 \eta(\frac{\tau}{2})^2\eta(12\tau)^2\eta(3\tau)^3}\ ,
\end{align}
from which the function $\phi_{\rm 12A}(\tau,z)$ can be reconstructed. However, it is easier to
analyse the modular properties of  $\phi_{{\rm 12A}}(\tau,z)$ using the
methods described in the next subsection.

\subsection{Studying the modular forms of weight two}\label{s:wei2}

Every weak Jacobi form of index $1$ can be written as a linear combination of the standard
Jacobi forms $\phi_{0,1}$ and $\phi_{-2,1}$ of weight $0$ and $-2$,  respectively
(see appendix \ref{App:Definitions}), where the
coefficients lie in the space of modular forms under the relevant subgroup of $\SL(2,\ZZ)$
\cite{EichlerZagier}.
In particular,
\be\label{phistandJac} \phi_g(\tau,z)=B_g\, \phi_{0,1}(\tau,z)+F_g(\tau)\, \phi_{-2,1}(\tau,z)\ ,
\ee
where $B_g$ is the constant
\be\label{ag} B_g=\frac{1}{12}\phi_g(\tau,0)=\frac{1}{24}\Tr_{\mathbf{23}\oplus\mathbf{1}}(g)\ ,
\ee
while $F_g$ is a suitable modular form of weight two. If $g$ is in \eqref{set1}, the phases in
\eqref{conj} are trivial, and $F_g\in M_2(\Gamma_0(N))$, where
$M_k(\Gamma)$ denotes the space of modular
form of weight $k$ under the group $\Gamma$. In \cite{Cheng}, it has been shown that, for all
$g$ in \eqref{set1}, there is a unique $F_g\in M_2(\Gamma_0(N))$  matching the known
coefficients of the $q$-expansion of $\phi_g$ (see also appendix \ref{a:weightwo}).

If $g$ is one of the classes in \eqref{set2}, then $B_g=0$ and  the conjecture \eqref{conj} would
imply that $F_g$ transforms as
\be\label{Fgtransf} F_g\Bigl(\frac{a\tau+b}{c\tau+d}\Bigr)=e^{\frac{2\pi i cd}{Nh}}(c\tau+d)^2F(\tau)\ ,\qquad \begin{pmatrix} a & b\\ c & d\end{pmatrix}\in\Gamma_0(N)\ ,
\ee
where $N$ is the order of $g$ and $h|\gcd(N,12)$. In particular, this means that
\be\label{cond2}
F_g\in  M_{2}(\Gamma_0(Nh))\ ,\qquad F_g^h\in M_{2h}(\Gamma_0(N))\ .
\ee
Conversely, if these conditions hold, then $F_g$ transforms as a modular form of weight $2$
under $\Gamma_0(N)$ up to a certain $h$-th root of unity $e^{\frac{2\pi i r cd}{Nh}}$, for some $r\in\ZZ$.

For all $g$ in  \eqref{set2}, we have found modular forms
$F_g\in M_2(\Gamma_0(Nh))$  for the above values of $h$
that reproduce the first few coefficients of $q$ as determined from (\ref{3.3}) and (\ref{3.4}),
and satisfy $F_g^h \in M_{2h}(\Gamma_0(N))$. The explicit expressions for $F_g$ (and $F_g^h$)
in terms of suitable bases of $M_2(\Gamma_0(Nh))$ (and $M_{2h}(\Gamma_0(N))$) are quite
complicated, and are therefore only given in appendix~\ref{a:weightwo}.

It remains to prove that the $h$-th root of unity $e^{\frac{2\pi i r cd}{Nh}}$ in the modular transformation of $F_g$ is the same as in \eqref{conj},
{\it i.e.}\ that $r=1$. We are mainly interested in the class 12A because the other
cases have already been dealt with in the previous subsection. For 12A we have found $h=2$,
and since the phase is non-trivial (as can be easily checked), the only possibility is $r=1$.
This completes our analysis.

%%%%%%%%%%%%%%%%%%%%%%%%%%%%%%%%%%%%%%%%%%%%%%%%%%%%%%%%%
\section{Decomposition into irreducible representations}\label{s:tables}

In the previous sections, we provided explicit expressions for all twining genera,
matching the first few coefficients in \eqref{phigH} and transforming as in \eqref{conj}.
The knowledge of all such characters also leads to another very stringent consistency check of
the proposal. It allows us to check
whether there are indeed underlying $\MM_{24}$ representations $H_n$ such
that $A_n(g)=\Tr_{H_n}(g)$, where $A_n(g)$ is the coefficient replacing $A_n$ in $\phi_g$, see
(\ref{phigH}).

In order to understand how this works, let us assume that $A_n(g)=\Tr_{H_n}(g)$, where
$H_n$ is a $\MM_{24}$ representation which we decompose as
\be\label{Hndecomp}
H_n=\bigoplus_{i=1}^{26}\, h_{n,i} \, R_i \ ,
\ee
where $i=1,\ldots,26$ labels the different irreducible representations of $\MM_{24}$, numbered
as in table \ref{t:char}. To start with we rewrite (\ref{phigH}) as
\be\label{4.2}
-\Tr_{H_0}(g)
+\sum_{n=1}^\infty\Tr_{H_n}(g)q^n =
q^{\frac{1}{8}}\frac{\eta(\tau)^3}{\vartheta_1(\tau,z)^2}\Bigl(2\phi_g(\tau,z)-
\Tr_{H_{00}}(g)\ch^{\N=4}_{h=\frac{1}{4},l=0}(\tau,z)\Bigr)\ ,
\ee
where we have used (\ref{long}). Next we recall that the characters of a finite group satisfy the
orthonormality relations
\be\label{chorth}
\sum_g c(g) \,\,\overline{\Tr_{R}(g)}\,\Tr_{R'}(g)=\begin{cases}1
&\text{if }R\cong R'\\0 &\text{otherwise,}\end{cases}
\ee
where $R$ and $R'$ are two irreducible representations of $\MM_{24}$. Here
the sum runs over all conjugacy classes of $\MM_{24}$, and
$c(g)^{-1}$ is the order of the centraliser of $g$%, which equals
\begin{equation}
c(g) = \frac{n(g)}{|\MM_{24}|} \ , \qquad \text{with} \quad
|\MM_{24}|=2^{10}\cdot 3^3\cdot 5\cdot 7\cdot 11\cdot 23
\end{equation}
the order of $\MM_{24}$ and $n(g)$ the number of elements in the conjugacy class of $g$.
Using (\ref{chorth}) we now obtain from (\ref{4.2})
\be\label{multipl}
-2 \delta_{i,1}+\sum_{n=1}^\infty h_{n,i}\, q^n=
q^{\frac{1}{8}}\frac{\eta(\tau)^3}{\vartheta_1(\tau,z)^2}
\Bigl(2 \sum_g c(g) \overline{\Tr_{R_i}(g)}\, \phi_g(\tau,z)
-(\delta_{i,1}+\delta_{i,2})\ch^{\N=4}_{h=\frac{1}{4},l=0}(\tau,z)\Bigr)\ ,
\ee
where $i=1,\ldots,26$. In deriving (\ref{multipl}) we have used that
$H_{00}=R_1\oplus R_2={\bf 1}\oplus {\bf 23}$ and
$H_0 = 2 \cdot R_1 = 2 \cdot {\bf 1}$. For $i>2$, \eqref{multipl} simplifies,
and we obtain
\begin{align}
\sum_{n=1}^\infty h_{n,i}\, q^n= -2 q^{\frac{1}{8}}\frac{\sum_g c(g) \overline{\Tr_{R_i}(g)}\,
F_g(\tau)}{\eta(\tau)^3}
=-2 \frac{\sum_g c(g) \overline{\Tr_{R_i}(g)}\,  F_g(\tau)}{\prod_{n=1}^\infty(1-q^n)^3}\ ,\qquad i>2\ ,
\end{align}
where we used \eqref{phistandJac}, as well as \eqref{ag} and  \eqref{chorth} again.
Furthermore, we have plugged in the explicit expression for $\phi_{-2,1}$ from (\ref{A.6}).

The character values $\Tr_{R_i}(g)$ are given in table~\ref{t:char},
and the explicit values of $c(g)$ are tabulated in table~\ref{t:cvalues}. The
analysis of the previous sections provide closed formulae for the twining characters
$\phi_g(\tau,z)$ for all conjugacy classes $g$ in ${\mathbb M}_{24}$, so that the right hand
side of \eqref{multipl} can be easily evaluated. Thus we can determine the multiplicities
$h_{n,i}$ explicitly. The statement that there are underlying $\MM_{24}$ representations
$H_n$  is now simply equivalent to the property of the multiplicities $h_{n,i}$ to be
non-negative integers. We have worked out these multiplicities for $n\leq 500$, and
all of them are indeed non-negative integers; the explicit values for $n\leq 30$ are listed in
table~\ref{t:decom}.

Since all the characters we have constructed have real coefficients,
the multiplicities of conjugate representations are always equal, so that all $H_n$ are real
representations, as expected. It is also remarkable that the multiplicities of the real irreducible
representations in \eqref{Hndecomp} are always even, at least, up to $n=500$ (see also
\cite{Cheng}). This suggests that the actual symmetry group may be slightly bigger than the
Mathieu group $\MM_{24}$.

%%%%%%%%%%%%%%%%%%%%%%%%%%%%%%%%%%%%%%%%%%%%%%%%%%%%%%%%%%%%%%%%%%%%%%%%%%%%%%%
\section{Conclusions}\label{Sect:Conclusions}

In this paper we have accumulated compelling evidence for the conjecture of
Eguchi, Ooguri and Tachikawa~\cite{EOT} that the states contributing to the elliptic genus of K3
carry an action of the Mathieu group $\MM_{24}$. More specifically, we have
found closed form expressions for the twining genera of K3 for all conjugacy classes
of $\MM_{24}$, thus completing the programme initiated in \cite{Cheng,Gaberdiel:2010ch}.
We have shown that the twining genera transform indeed as Jacobi forms of index one and weight
zero  under $\Gamma_{0}(N)$, where $N$ is the order of the corresponding group element. The
twining genera of the conjugacy classes that have no representative contained in
$\MM_{23}\subset\MM_{24}$ have a non-trivial multiplier system, which we have identified, see
eq.~(\ref{conj}).

The explicit knowledge of all twining genera allows one to determine the decomposition of the
elliptic genus of K3 in terms of $\MM_{24}$ representations, and we have checked that
the multiplicities with which these representations appear are indeed non-negative integers, at least
for the first 500 coefficients --- see also table~\ref{t:decom} for explicit results for $n\leq 30$. This is a highly
non-trivial consistency check;  indeed, if we were able to show that {\em all} of these multiplicities are
non-negative integers, this would effectively prove the conjecture of \cite{EOT}.

Another, more conceptual, proof of the conjecture would consist of constructing explicitly the action
of $\MM_{24}$ on the BPS states contributing to the elliptic genus of K3.
For some generators in $\MM_{23}\subset\MM_{24}$ this can be done fairly directly since
they describe geometric automorphisms of the K3 surface \cite{Mukai,Kondo}, see
also \cite{Taormina:2010pf} for some recent progress in this direction. However,
it seems unlikely that these purely geometrical symmetries will suffice to account for the full
$\MM_{24}$ symmetry of the elliptic genus. In fact, one may guess that one will need to
consider the full moduli space of the non-linear sigma model in order to achieve this. Thus at least
some of the required symmetries may have an interpretation in terms of truly stringy symmetries,
such as {\it e.g.}~T-duality. It would be very interesting to explore these ideas in more detail \cite{pro}.
\smallskip

On a more technical note, it is amusing to note (see also \cite{Gaberdiel:2010ch})  that, in contrast
to the McKay-Thompson series in Monstrous Moonshine~\cite{CN}, not all twining genera
seem to be Hauptmoduls for genus zero congruence subgroups (although many are). For example, the
twining character 21AB,  for which the relevant modular group, $\Gamma_0(21)$, is not genus 0. This
fact is also reflected in equation (\ref{formPhi}) where we have given the explicit expression of
$\Phi_{{\rm 21AB}}^{(3)}(\tau)$ as a modular form of $\Gamma_0(21)$. Unlike the remaining
twining characters, it cannot be written as a rational function of a single modular function
(the Hauptmodul) but rather involves two functions (the McKay-Thompson series $T_{[\rm 21B]}$ and
$T_{[\rm 21D]}$). It would be very interesting to understand what the significance of the genus
zero property in this context is.

%%%%%%%%%%%%%%%%%%%%%%%%%%%%%%%%%%%%%%%%%%%%%%%%%%%%%%%%%%%%%%%%%%%%%%%%%%%%%%%
\section*{Acknowledgments}
We thank Terry Gannon for useful communications and John McKay for inspiring discussions.
The research of MRG and SH is partially
supported by a grant from the Swiss National Science Foundation, and the research of RV
is supported by an INFN Fellowship.
\bigskip

%\section{Tables}\label{s:tables}

\begin{table}[H]\centerline{
$\begin{array}{|c|c||c|c||c|c|}\hline
&&&&&\\[-12pt]
\text{Class}  & c(g)^{-1} & \text{Class}  & c(g)^{-1} & \text{Class}  & c(g)^{-1} \\
&&&&&\\[-12pt]\hline
&&&&&\\[-10pt]
{\rm 1A} & 244 823 040 & {\rm  6A} & 24 & {\rm  4C} & 96\\
&&&&&\\[-10pt]
{\rm  2A} & 21 504 & {\rm  11A} & 11 & {\rm 3B} & 504\\
&&&&&\\[-10pt]
{\rm  3A} & 1 080 & {\rm 15AB} & 15 & {\rm  2B} & 7 680\\
&&&&&\\[-10pt]
{\rm 5A} & 60 & {\rm 14AB} & 14 & {\rm 10A} & 20\\
&&&&&\\[-10pt]
{\rm 4B} & 128 & {\rm 23AB} & 23 & {\rm 21AB} & 21\\
&&&&&\\[-10pt]
{\rm 7AB} & 42 & {\rm 12B} & 12 & {\rm 4A} & 384\\
&&&&&\\[-10pt]
{\rm 8A} & 16 & {\rm 6B} & 24 & {\rm 12A} & 12\\[2pt]\hline
\end{array}$}\caption{Order $c(g)^{-1}$ of the centraliser of the class $g$.}\label{t:cvalues}
\end{table}

\newpage

\begin{table}[H]\centerline{\scalebox{.95}{
\rotatebox{90}{$\begin{array}{r|rrrrrrrrrrrrrrrrrrrrrrrrrrrrrrrrrrrrrr}
 %&1 & 2 & 3 & 4 & 5 & 6 & 7 & 8 & 9 & 10 & 11 & 12 & 13 & 14 & 15 & 16 & 17 & 18 & 19 & 20 & 21 & 22 & 23 & 24 & 25 & 26\\\hline
 i&{\rm 1A}  &  {\rm 2A} &  {\rm 3A} & {\rm  5A} &  {\rm 4B} &  {\rm 7A}
 &  {\rm 7B} &  {\rm 8A} &  {\rm 6A} &  {\rm 11A} &  {\rm15A} &  {\rm 15B}
 &  {\rm 14A} &  {\rm 14B} &  {\rm 23A}  &  {\rm 23B} &  {\rm 12B} &  {\rm 6B}
 &  {\rm 4C} &  {\rm 3B} &  {\rm 2B} &  {\rm 10A} &  {\rm 21A} &  {\rm 21B} &  {\rm 4A} &  {\rm 12A}\\\hline
% &    &   &   &  {\rm 2A} &   &
% &  {\rm 4B} &  {\rm 3A} &   &  {\rm 3A} &  {\rm 3A} &  {\rm 2A} &  {\rm 2A} &
% &  &   {\rm 6B} &  {\rm 3B} &  {\rm 2B} &   &   &  {\rm 2B} &  {\rm 3B} &  {\rm 3B} &  {\rm 2A} &  {\rm 6A}\\
%   &   &   &   &   &   &   &   &  {\rm 2A} &  &  &  &  &  &  &  &  {\rm 4C} &  {\rm 2B} &   &   &   &   &   &   &   &  {\rm 3A}\\\hline
 1& 1 & 1 & 1 & 1 & 1 & 1 & 1 & 1 & 1 & 1 & 1 & 1 & 1 & 1 & 1 & 1 & 1 & 1 & 1 & 1 & 1 & 1 & 1 & 1 & 1 & 1 \\
 2&23 & 7 & 5 & 3 & 3 & 2 & 2 & 1 & 1 & 1 & 0 & 0 & 0 & 0 & 0 & 0 & -1 & -1 & -1 & -1 & -1 & -1 & -1 & -1 & -1 &
   -1 \\
 3 & 252 & 28 & 9 & 2 & 4 & 0 & 0 & 0 & 1 & -1 & -1 & -1 & 0 & 0 & -1 & -1 & 0 & 0 & 0 & 0 & 12 & 2 & 0 & 0 & 4 & 1
   \\
 4 &253 & 13 & 10 & 3 & 1 & 1 & 1 & -1 & -2 & 0 & 0 & 0 & -1 & -1 & 0 & 0 & 1 & 1 & 1 & 1 & -11 & -1 & 1 & 1 & -3 &
   0 \\
 5 &1771 & -21 & 16 & 1 & -5 & 0 & 0 & -1 & 0 & 0 & 1 & 1 & 0 & 0 & 0 & 0 & -1 & -1 & -1 & 7 & 11 & 1 & 0 & 0 & 3 &
   0 \\
 6 &3520 & 64 & 10 & 0 & 0 & -1 & -1 & 0 & -2 & 0 & 0 & 0 & 1 & 1 & 1 & 1 & 0 & 0 & 0 & -8 & 0 & 0 & -1 & -1 & 0 &
   0 \\
 7& 45 & -3 & 0 & 0 & 1 & e_7^+ & e_7^- & -1 & 0 & 1 & 0 & 0 & -e_7^+ & -e_7^- & -1 & -1 & 1 &
   -1 & 1 & 3 & 5 & 0 & e_7^- & e_7^+ & -3 & 0 \\
 8 &45 & -3 & 0 & 0 & 1 & e_7^- & e_7^+ & -1 & 0 & 1 & 0 & 0 & -e_7^- & -e_7^+ & -1 & -1 & 1 &
   -1 & 1 & 3 & 5 & 0 & e_7^+ & e_7^- & -3 & 0 \\
 9& 990 & -18 & 0 & 0 & 2 & e_7^+ & e_7^- & 0 & 0 & 0 & 0 & 0 & e_7^+ & e_7^- & 1 & 1 & 1 & -1
   & -2 & 3 & -10 & 0 & e_7^- & e_7^+ & 6 & 0 \\
 10&990 & -18 & 0 & 0 & 2 & e_7^- & e_7^+ & 0 & 0 & 0 & 0 & 0 & e_7^- & e_7^+ & 1 & 1 & 1 & -1
   & -2 & 3 & -10 & 0 & e_7^+ & e_7^- & 6 & 0 \\
 11&1035 & -21 & 0 & 0 & 3 & 2 e_7^+ & 2 e_7^- & -1 & 0 & 1 & 0 & 0 & 0 & 0 & 0 & 0 & -1 & 1 & -1 & -3 &
   -5 & 0 & -e_7^- & -e_7^+ & 3 & 0 \\
 12&1035 & -21 & 0 & 0 & 3 & 2 e_7^- & 2 e_7^+ & -1 & 0 & 1 & 0 & 0 & 0 & 0 & 0 & 0 & -1 & 1 & -1 & -3 &
   -5 & 0 & -e_7^+ & -e_7^- & 3 & 0 \\
 13&1035 & 27 & 0 & 0 & -1 & -1 & -1 & 1 & 0 & 1 & 0 & 0 & -1 & -1 & 0 & 0 & 0 & 2 & 3 & 6 & 35 & 0 & -1 & -1 & 3 &
   0 \\
 14&231 & 7 & -3 & 1 & -1 & 0 & 0 & -1 & 1 & 0 & e_{15}^+ & e_{15}^- & 0 & 0 & 1 & 1 & 0 & 0 & 3 & 0 & -9 & 1 &
   0 & 0 & -1 & -1 \\
 15&231 & 7 & -3 & 1 & -1 & 0 & 0 & -1 & 1 & 0 & e_{15}^- & e_{15}^+ & 0 & 0 & 1 & 1 & 0 & 0 & 3 & 0 & -9 & 1 &
   0 & 0 & -1 & -1 \\
 16&770 & -14 & 5 & 0 & -2 & 0 & 0 & 0 & 1 & 0 & 0 & 0 & 0 & 0 & e_{23}^+ & e_{23}^- & 1 & 1 & -2 & -7 & 10 & 0
   & 0 & 0 & 2 & -1 \\
 17&770 & -14 & 5 & 0 & -2 & 0 & 0 & 0 & 1 & 0 & 0 & 0 & 0 & 0 & e_{23}^- & e_{23}^+ & 1 & 1 & -2 & -7 & 10 & 0
   & 0 & 0 & 2 & -1 \\
 18&483 & 35 & 6 & -2 & 3 & 0 & 0 & -1 & 2 & -1 & 1 & 1 & 0 & 0 & 0 & 0 & 0 & 0 & 3 & 0 & 3 & -2 & 0 & 0 & 3 & 0 \\
 19&1265 & 49 & 5 & 0 & 1 & -2 & -2 & 1 & 1 & 0 & 0 & 0 & 0 & 0 & 0 & 0 & 0 & 0 & -3 & 8 & -15 & 0 & 1 & 1 & -7 &
   -1 \\
 20&2024 & 8 & -1 & -1 & 0 & 1 & 1 & 0 & -1 & 0 & -1 & -1 & 1 & 1 & 0 & 0 & 0 & 0 & 0 & 8 & 24 & -1 & 1 & 1 & 8 &
   -1 \\
 21&2277 & 21 & 0 & -3 & 1 & 2 & 2 & -1 & 0 & 0 & 0 & 0 & 0 & 0 & 0 & 0 & 0 & 2 & -3 & 6 & -19 & 1 & -1 & -1 & -3 &
   0 \\
 22&3312 & 48 & 0 & -3 & 0 & 1 & 1 & 0 & 0 & 1 & 0 & 0 & -1 & -1 & 0 & 0 & 0 & -2 & 0 & -6 & 16 & 1 & 1 & 1 & 0 & 0
   \\
 23&5313 & 49 & -15 & 3 & -3 & 0 & 0 & -1 & 1 & 0 & 0 & 0 & 0 & 0 & 0 & 0 & 0 & 0 & -3 & 0 & 9 & -1 & 0 & 0 & 1 & 1
   \\
 24&5796 & -28 & -9 & 1 & 4 & 0 & 0 & 0 & -1 & -1 & 1 & 1 & 0 & 0 & 0 & 0 & 0 & 0 & 0 & 0 & 36 & 1 & 0 & 0 & -4 &
   -1 \\
 25&5544 & -56 & 9 & -1 & 0 & 0 & 0 & 0 & 1 & 0 & -1 & -1 & 0 & 0 & 1 & 1 & 0 & 0 & 0 & 0 & 24 & -1 & 0 & 0 & -8 &
   1 \\
 26&10395 & -21 & 0 & 0 & -1 & 0 & 0 & 1 & 0 & 0 & 0 & 0 & 0 & 0 & -1 & -1 & 0 & 0 & 3 & 0 & -45 & 0 & 0 & 0 & 3 &
   0
\end{array}$}}}\caption{The character table of the Mathieu group $\MM_{24}$. The rows correspond to the
representations $R_i$, numbered from $1$ to $26$, while the columns describe the different conjugacy classes.
Finally, $e_p^\pm=(-1\pm i\sqrt{p})/2$.}\label{t:char}
\end{table}

\begin{table}[H]\centerline{\scalebox{.85}{
\rotatebox{90}{$\begin{array}{|c|ccccccccccccccccccccc|}\hline
 i& 1 & 2 & 3 & 4 & 5 & 6 & 7,8 & 9,10 & 11,12 & 13 & 14,15 & 16,17 & 18 & 19 & 20 & 21 & 22 & 23 & 24 & 25 & 26 \\\hline
 n& \mathbf{1} & \mathbf{23} & \mathbf{252} & \mathbf{ 253} & \mathbf{ 1771} & \mathbf{ 3520} & \mathbf{ 45} & \mathbf{ 990} & \mathbf{ 1035} & \mathbf{ 1035} & \mathbf{ 231} & \mathbf{ 770} & \mathbf{ 483} & \mathbf{ \
 1265} & \mathbf{ 2024} & \mathbf{ 2277} & \mathbf{ 3312} & \mathbf{ 5313} & \mathbf{ 5796} & \mathbf{ 5544} & \mathbf{ 10395}\\ \hline
 1 & 0 & 0 & 0 & 0 & 0 & 0 & 1 & 0 & 0 & 0 & 0 & 0 & 0 & 0 & 0 & 0 & 0
   & 0 & 0 & 0 & 0 \\
 2 & 0 & 0 & 0 & 0 & 0 & 0 & 0 & 0 & 0 & 0 & 1 & 0 & 0 & 0 & 0 & 0 & 0
   & 0 & 0 & 0 & 0 \\
 3 & 0 & 0 & 0 & 0 & 0 & 0 & 0 & 0 & 0 & 0 & 0 & 1 & 0 & 0 & 0 & 0 & 0
   & 0 & 0 & 0 & 0 \\
 4 & 0 & 0 & 0 & 0 & 0 & 0 & 0 & 0 & 0 & 0 & 0 & 0 & 0 & 0 & 0 & 2 & 0
   & 0 & 0 & 0 & 0 \\
 5 & 0 & 0 & 0 & 0 & 0 & 0 & 0 & 0 & 0 & 0 & 0 & 0 & 0 & 0 & 0 & 0 & 0
   & 0 & 2 & 0 & 0 \\
 6 & 0 & 0 & 0 & 0 & 0 & 2 & 0 & 0 & 0 & 0 & 0 & 0 & 0 & 0 & 0 & 0 & 0
   & 0 & 0 & 0 & 2 \\
 7 & 0 & 0 & 0 & 0 & 2 & 0 & 0 & 0 & 0 & 0 & 0 & 0 & 0 & 0 & 2 & 0 & 0
   & 2 & 2 & 2 & 2 \\
 8 & 0 & 0 & 0 & 0 & 0 & 2 & 0 & 1 & 1 & 0 & 0 & 0 & 0 & 2 & 0 & 2 & 2
   & 4 & 2 & 2 & 6 \\
 9 & 0 & 0 & 0 & 0 & 2 & 4 & 0 & 0 & 2 & 2 & 0 & 2 & 2 & 0 & 2 & 2 & 4
   & 4 & 8 & 8 & 10 \\
 10 & 0 & 0 & 0 & 2 & 4 & 8 & 0 & 2 & 2 & 2 & 2 & 0 & 2 & 4 & 4 & 6 &
   6 & 12 & 10 & 10 & 24 \\
 11 & 0 & 0 & 0 & 0 & 8 & 12 & 0 & 4 & 4 & 6 & 0 & 4 & 0 & 2 & 10 & 8
   & 14 & 22 & 26 & 24 & 40 \\
 12 & 0 & 2 & 2 & 4 & 12 & 30 & 0 & 8 & 8 & 4 & 2 & 6 & 4 & 12 & 12 &
   18 & 26 & 40 & 40 & 38 & 80 \\
 13 & 0 & 0 & 4 & 2 & 26 & 44 & 2 & 14 & 14 & 18 & 2 & 10 & 6 & 16 &
   30 & 28 & 44 & 70 & 84 & 80 & 136 \\
 14 & 0 & 0 & 4 & 6 & 38 & 86 & 0 & 24 & 24 & 22 & 8 & 16 & 14 & 34 &
   46 & 58 & 80 & 128 & 132 & 126 & 254 \\
 15 & 0 & 0 & 12 & 8 & 78 & 144 & 2 & 40 & 44 & 46 & 8 & 38 & 18 & 46
   & 86 & 88 & 138 & 218 & 246 & 238 & 424 \\
 16 & 0 & 2 & 18 & 22 & 122 & 252 & 2 & 72 & 72 & 68 & 18 & 50 & 36 &
   100 & 140 & 170 & 232 & 378 & 400 & 382 & 742 \\
 17 & 0 & 2 & 30 & 26 & 212 & 410 & 8 & 116 & 124 & 130 & 25 & 94 & 54
   & 140 & 246 & 262 & 392 & 630 & 704 & 670 & 1222 \\
 18 & 0 & 6 & 50 & 58 & 342 & 704 & 6 & 194 & 202 & 192 & 50 & 148 &
   100 & 256 & 388 & 454 & 654 & 1044 & 1120 & 1074 & 2058 \\
 19 & 0 & 4 & 80 & 72 & 582 & 1116 & 18 & 318 & 332 & 346 & 68 & 252 &
   150 & 394 & 664 & 722 & 1062 & 1702 & 1880 & 1800 & 3320 \\
 20 & 0 & 14 & 128 & 138 & 904 & 1836 & 20 & 516 & 536 & 520 & 126 &
   390 & 254 & 676 & 1036 & 1196 & 1716 & 2764 & 2980 & 2846 & 5408 \\
 21 & 2 & 20 & 214 & 200 & 1476 & 2902 & 40 & 814 & 860 & 872 & 182 &
   652 & 396 & 1020 & 1684 & 1862 & 2742 & 4384 & 4828 & 4622 & 8572
   \\
 22 & 2 & 32 & 328 & 346 & 2302 & 4616 & 55 & 1298 & 1348 & 1336 & 314
   & 988 & 640 & 1686 & 2630 & 3000 & 4324 & 6950 & 7532 & 7204 &
   13620 \\
 23 & 2 & 40 & 512 & 496 & 3638 & 7166 & 98 & 2020 & 2118 & 2144 & 460
   & 1590 & 972 & 2546 & 4162 & 4624 & 6768 & 10856 & 11898 & 11376 &
   21204 \\
 24 & 0 & 80 & 798 & 824 & 5584 & 11192 & 132 & 3140 & 3278 & 3236 &
   744 & 2426 & 1544 & 4050 & 6376 & 7248 & 10500 & 16834 & 18294 &
   17504 & 32976 \\
 25 & 8 & 108 & 1232 & 1208 & 8654 & 17084 & 234 & 4814 & 5038 & 5084
   & 1106 & 3764 & 2336 & 6108 & 9892 & 11042 & 16112 & 25840 & 28288
   & 27056 & 50524 \\
 26 & 6 & 174 & 1860 & 1904 & 13090 & 26148 & 322 & 7348 & 7670 & 7626
   & 1742 & 5677 & 3602 & 9444 & 14968 & 16940 & 24566 & 39428 & 42894
   & 41022 & 77176 \\
 27 & 12 & 252 & 2836 & 2802 & 19914 & 39436 & 514 & 11092 & 11618 &
   11666 & 2560 & 8688 & 5394 & 14100 & 22744 & 25462 & 37148 & 59564
   & 65114 & 62294 & 116494 \\
 28 & 16 & 398 & 4238 & 4310 & 29772 & 59330 & 742 & 16686 & 17418 &
   17356 & 3922 & 12912 & 8160 & 21414 & 34026 & 38434 & 55764 & 89490
   & 97456 & 93218 & 175146 \\
 29 & 26 & 560 & 6328 & 6286 & 44512 & 88280 & 1154 & 24840 & 25994 &
   26078 & 5758 & 19380 & 12090 & 31636 & 50892 & 57068 & 83146 &
   133356 & 145690 & 139342 & 260828 \\
 30 & 34 & 876 & 9368 & 9486 & 65776 & 131020 & 1642 & 36824 & 38480 &
   38368 & 8642 & 28580 & 18008 & 47172 & 75158 & 84776 & 123176 &
   197596 & 215318 & 205970 & 386724
\end{array}$}}}\caption{Multiplicities $h_{n,i}$ in the decomposition of the representations $H_n=\oplus_{i=1}^{26} h_{n,i}R_i$, for $n\le 30$. The representations $R_i$ are numbered as in table~3, in the second row the dimension is given. Pairs of conjugate representations are listed together, because they always appear with the same multiplicities.}\label{t:decom}
\end{table}

\newpage

\appendix

\section{Definitions}\label{App:Definitions}
Our conventions for  the Dedekind eta and the
Jacobi theta functions are
\begin{align}
\eta(\tau) & =  q^{\frac{1}{24}} \prod_{n=1}^{\infty} (1 - q^n) \nonumber \\
\vartheta_1(\tau,z) & = -iq^{\frac{1}{8}} y^{\frac{1}{2}}\, \prod_{n=1}^\infty(1-q^n)(1-yq^n)(1-y^{-1}q^{n-1})
\nonumber \\
\vartheta_2(\tau,z) & = 2\, q^{\frac{1}{8}} \cos(\pi z)\, \prod_{n=1}^{\infty} (1-q^n)\, (1+yq^n)
(1+y^{-1} q^n)   \nonumber \\
\vartheta_3(\tau,z) & =    \prod_{n=1}^{\infty} (1-q^n) \, (1+yq^{n-1/2})(1+y^{-1}q^{n-1/2}) \\
\vartheta_4(\tau,z) & =    \prod_{n=1}^{\infty} (1-q^n) \, (1-yq^{n-1/2}) (1-y^{-1}q^{n-1/2}) \ \ .\nonumber
\end{align}
Under modular transformations the $\vartheta$  and $\eta$ functions transform as
\begin{align}
&\vartheta_1(\tau+1,z) = e^{ \frac{2\pi i}{8}} \, \vartheta_1(\tau,z)
& &\vartheta_1(- \tfrac{1}{\tau},\tfrac{z}{\tau}) = -(-i \tau)^{\frac{1}{2}}\,
e^{\frac{i\pi z^2}{\tau}}\,  \vartheta_1(\tau,z) \ , \\
&\vartheta_2(\tau+1,z) = e^{ \frac{2\pi i}{8}} \, \vartheta_2(\tau,z)
& &\vartheta_2(- \tfrac{1}{\tau},\tfrac{z}{\tau}) = (-i \tau)^{\frac{1}{2}}\,
e^{\frac{i\pi z^2}{\tau}}\,  \vartheta_4(\tau,z) \ ,\\
&\vartheta_3(\tau+1,z) = \vartheta_4(\tau,z)
& &\vartheta_3(- \tfrac{1}{\tau},\tfrac{z}{\tau}) =(-i \tau)^{\frac{1}{2}}\,
e^{\frac{i\pi z^2}{\tau}}\,  \vartheta_3(\tau,z) \ , \\
&\vartheta_4(\tau+1,z) = \vartheta_3(\tau,z)
& &\vartheta_4(- \tfrac{1}{\tau},\tfrac{z}{\tau}) = (-i \tau)^{\frac{1}{2}}\,
e^{\frac{i\pi z^2}{\tau}}\,  \vartheta_2(\tau,z) \ ,
\end{align}
as well as
\begin{align}
&\eta(\tau+1) =  e^{ \frac{2\pi i}{24}} \, \eta(\tau) \quad
&&\eta(-\tfrac{1}{\tau}) = (-i \tau)^{\frac{1}{2}}\, \eta(\tau) \ .
\end{align}
The theta constants $\vartheta_a(\tau)$ are defined as $\vartheta_a(\tau)\equiv \vartheta_a(\tau,z=0)$.
The standard weak Jacobi forms $\phi_{0,1}$ and $\phi_{-2,1}$ of index $1$ and weight $0$ and $2$ can be defined as \cite{EichlerZagier}
\be \label{A.6}
\phi_{0,1}(\tau,z)=4\sum_{i=2}^4\frac{\vartheta_i(\tau,z)^2}{\vartheta_i(\tau,0)^2}\ ,\qquad \qquad \phi_{-2,1}(\tau,z)=-\frac{\vartheta_1(\tau,z)^2}{\eta(\tau)^6}\ .
\ee

For completeness we give here the first few terms of the McKay-Thompson series that
appear in our analysis
\begin{align}
&T_{\rm [8E]}=\frac{1}{q}+4q+2q^3-8q^5-q^7+20q^9-2q^{11}-40q^{13}+\cdots\,,\\
&T_{\rm [10E]}=\frac{1}{q}+q+2 q^2+2 q^3-2 q^4-q^5-4 q^7-2 q^8+5 q^9+2 q^{10}+8 q^{12}+\cdots\,,\\
&T_{\rm [12I]}=\frac{1}{q}+2 q+q^3-2 q^7-2 q^9+2 q^{11}+4 q^{13}+3 q^{15}+\cdots\,,\\
&T_{\rm [21B]}=\frac{1}{q}-q-q^2+q^3+2q^4-q^5+3q^6-q^7-q^8-2q^9+q^{11}+\cdots\,,\\
&T_{\rm [21D]}=\frac{1}{q}+5q+8q^2+16q^3+26q^4+44q^5+66q^6+104q^7+\cdots\,,
\end{align}
see  \cite{FMN,webInteger} for more information about these series.

\section{Modular forms for $\Gamma_0(N)$}\label{a:modforms}

In this section, we will describe a basis of the space $M_k(\Gamma_0(N))$ of modular
forms of weight $k$ under $\Gamma_0(N)$. Our main references for this section are
\cite{Lang,Stein}.

The space $M_k(\Gamma_0(N))$ of modular forms of weight $k$ splits into a direct sum
\be\label{modspacedec} M_k(\Gamma_0(N))=E_k(\Gamma_0(N))\oplus S_k(\Gamma_0(N))\ ,
\ee
where $S_k(\Gamma_0(N))$ is the space of cusp forms, {\it i.e.}\ forms they vanish at all the
cusps\footnote{The cusps correspond to $\Gamma_0(N)$-orbits in $\QQ\cup\{\infty\}$,
with $\Gamma_0(N)$ acting by fractional linear transformations.} of
$\overline{\HH/\Gamma_0(N)}$. The space $E_k(\Gamma_0(N))$ is defined as the unique
subspace satisfying \eqref{modspacedec} that is invariant under the action of all Hecke operators
\cite{Stein}.
\smallskip

A convenient basis for $E_k(\Gamma_0(N))$ is given by (generalised) Eisenstein series \cite{Miyake}.
Let
\be E_k(\tau)=-\frac{B_k}{2k}+\sum_{n=1}^\infty \Bigl(\sum_{d|n}d^{k-1}\Bigr)q^n
\ee
be the standard Eisenstein series of weight $k$,
where $B_k$ are the Bernoulli numbers.
For $k>2$, $k$ even, $E_k$ is a modular form of weight $k$ under $SL(2,\ZZ)$, whereas
\be E_2\Bigl(\frac{a\tau+b}{c\tau+d}\Bigr)=(c\tau+d)^2E_2(\tau)-\frac{1}{4\pi i}c(c\tau+d)\ .
\ee
The definition of the Eisenstein series can be generalised to include modular forms
under $\Gamma_0(N)$. In particular,\footnote{The Eisenstein series $\psi^{(N)}$ are
related to the modular forms $\phi^{(N)}$ in \cite{Cheng} by $\psi^{(N)}=\frac{N-1}{24}\phi^{(N)}(\tau)$.}
\be \psi^{(N)}=q\frac{\partial}{\partial q}\log\frac{\eta(N\tau)}{\eta(\tau)}=E_2(\tau)- N E_2(N\tau)
\ee
is a modular form of weight $2$ under $\Gamma_0(N)$.
%Furthermore, let $\chi_m$ be a Dirichlet character of modulus $m$.
%, i.e. a map $\chi_m:\ZZ\to\CC$ such that
%\begin{enumerate}
%\item $\chi_m(a+m)=\chi(a)$,
%\item $\chi_m(a)=0$ if $(a,m)> 1$,
%\item $\chi_m(ab)=\chi(a)\chi(b)$.
%\end{enumerate}
The (generalised) Eisenstein series
\be E_k^{\chi_m}(\tau)=\sum_{n=1}^\infty \Bigl(\sum_{d|n}\,\overline{\chi_m(d)}\,\chi_m(n/d)\,d^{k-1}\Bigr)\,
q^n \ ,
\ee
where $k$ is even and $\chi_m$ is a non-trivial Dirichlet character of modulus $m$,
is a modular form of weight $k$ under $\Gamma_0(m^2)$. The only cases we need are
\be E_k^{(9)}=E_k^{\chi_3}\in M_k(\Gamma_0(9))\ ,\quad
E_k^{(16)}=E_k^{\chi_4}\in M_k(\Gamma_0(16))\ ,\quad
E_k^{(144)}=E_k^{\chi_{12}}\in M_k(\Gamma_0(144))\ ,
\ee
where $\chi_3$, $\chi_4$ and $\chi_{12}$ are the primitive Dirichlet characters of modulus
$3$, $4$ and $12$, that are uniquely determined by
\be \chi_3(2)=-1\ ,\qquad \chi_4(3)=-1\ ,\qquad \chi_{12}(5)=\chi_{12}(7)=-1\ .
\ee
\smallskip

In general, the space $S_k(\Gamma_0(N))$ is not generated by Eisenstein series.
It is obvious that if $M|N$, then $M_k(\Gamma_0(M))\subset M_k(\Gamma_0(N))$.
More generally, it is easy to see that if $f\in M_k(\Gamma_0(M))$, then, for any
divisor $n$ of $N/M$, $f(n\tau)\in\Gamma_0(N)$. For all $M|N$ and $n|(N/M)$, we
define the map
\begin{align}
\alpha_n:
&\ M_k(\Gamma_0(M))\to M_k(\Gamma_0(N))\\
&\ f(\tau)\mapsto \alpha_n(f)(\tau)=f(n\tau)\ .
\end{align}
The map $\alpha_n$ sends cusp forms to cusp forms and the union of the images
$\alpha_n(S_k(\Gamma_0(N/n)))\subseteq S_k(\Gamma_0(N))$, for all $n|N$, $n>1$,
is called the \emph{old subspace} of cusp forms. The complement of the old subspace
which is invariant under all Hecke operators is called the \emph{new subspace}. Thus,
we have a decomposition
\be S_k(\Gamma_0(N))=\bigoplus_{M|N}\bigoplus_{n|(N/M)}\alpha_n(S_k(\Gamma_0(M))_{new})\ ,
\ee
where $S_k(\Gamma_0(M))_{new}$ is the new subspace for $\Gamma_0(M)$. A basis for
$S_k(\Gamma_0(M))_{new}$ for the cases of interest are listed below;
a more extended list of their coefficients can be found at {\tt  http://modi.countnumber.de/}.
Some Fourier expansions have been computed using {\sf SAGE} ({\tt http://www.sagemath.org/}).
\begin{itemize}
\item Cusp forms $f_M\in S_2(\Gamma_0(M))_{new}$ --- if there is more than one generator,
the different generators are denoted by $f_{M,a}, f_{M,b}, \ldots$.
\begin{align}
 f_{21}(\tau)=&q-q^2+q^3-q^4-2 q^5-q^6-q^7+3 q^8+q^9+2 q^{10}+\cdots\\
 f_{23,a}(\tau)=&q-q^3-q^4-2 q^6+2 q^7-q^8+2 q^9+2 q^{10}+\cdots\\
 f_{23,b}(\tau)=&-q^2+2 q^3+q^4-2 q^5-q^6-2 q^7+2 q^8+2 q^{10}+\cdots\\
 f_{24}(\tau)=&q-q^3-2 q^5+q^9+4 q^{11}-2 q^{13}+2 q^{15}+2 q^{17}+\cdots\\
 f_{48}(\tau)=&q+q^3-2 q^5+q^9-4 q^{11}-2 q^{13}-2 q^{15}+2 q^{17}+\cdots\\
 f_{63,a}(\tau)=&q+q^2-q^4+2 q^5-q^7-3 q^8+2 q^{10}-4 q^{11}-2 q^{13}+\cdots\\
 f_{63,b}(\tau)=&q+q^4+q^7-6 q^{10}+2 q^{13}-5 q^{16}+\cdots\\
 f_{63,c}(\tau)=&q^2-2 q^5-q^8+2 q^{11}+q^{14}+2 q^{17}+\cdots\\
 f_{72}(\tau)=&q+2 q^5-4 q^{11}-2 q^{13}-2 q^{17}-4 q^{19}+\cdots\\
 f_{144,a}(\tau)=&q+4 q^7+2 q^{13}-8 q^{19}-5 q^{25}+4 q^{31}+\cdots\\
 f_{144,b}(\tau)=&q+2 q^5+4 q^{11}-2 q^{13}-2 q^{17}+4 q^{19}+\cdots\ .
 \end{align}
\item Cusp forms $g_M\in S_4(\Gamma_0(M))_{new}$
\begin{align}
 &g_5(\tau)=q-4 q^2+2 q^3+8 q^4-5 q^5-8 q^6+6 q^7-23 q^9+\cdots\\
 &g_6(\tau)=q-2 q^2-3 q^3+4 q^4+6 q^5+6 q^6-16 q^7-8 q^8+9 q^9+\cdots\\
 &g_8(\tau)=q-4 q^3-2 q^5+24 q^7-11 q^9-44 q^{11}+22 q^{13}+\cdots\\
 &g_{10}(\tau)=q+2 q^2-8 q^3+4 q^4+5 q^5-16 q^6-4 q^7+8 q^8+37 q^9+\cdots\\
 &g_{12}(\tau)=q+3 q^3-18 q^5+8 q^7+9 q^9+36 q^{11}-10 q^{13}-54 q^{15}+\cdots\ .
\end{align}
\item Cusp forms $h_M\in S_6(\Gamma_0(M))_{new}$
\begin{align}
h_3(\tau)=&q-6 q^2+9 q^3+4 q^4+6 q^5-54 q^6-40 q^7+168 q^8+\cdots \\
h_{7,a}(\tau)=&q-10 q^2-14 q^3+68 q^4-56 q^5+140 q^6-49 q^7-360 q^8+\cdots \\
h_{7,b}(\tau)=&q+4 q^2-2 q^4-14 q^5-84 q^6+49 q^7-10 q^8+\cdots\\
h_{7,c}(\tau)=&q^2-6 q^3+9 q^4+10 q^5-30 q^6+11 q^8+\cdots \\
h_{21,a}(\tau)=&q+q^2-9 q^3-31 q^4-34 q^5-9 q^6-49 q^7-63 q^8+\cdots\\
h_{21,b}(\tau)=&q+5 q^2+9 q^3-7 q^4+94 q^5+45 q^6-49 q^7-195 q^8+\cdots\\
h_{21,c}(\tau)=&q - 6 q^2 - 9 q^3 + 4 q^4 + 78 q^5 + 54 q^6 + 49 q^7 + 168 q^8 +
\cdots \\
h_{21,d}(\tau)=&q+10 q^2+9 q^3+68 q^4-106 q^5+90 q^6-49 q^7+360 q^8+\cdots\ .
\end{align}
\item Cusp forms $k_M\in S_8(\Gamma_0(M))_{new}$
\begin{align}
k_2(\tau)=q-8 q^2+12 q^3+64 q^4-210 q^5-96 q^6+1016 q^7-512 q^8+\cdots\ .
\end{align}
\item Cusp forms $l_M\in S_{12}(\Gamma_0(M))_{new}$
\begin{align}
l_3(\tau)=&q+78 q^2-243 q^3+4036 q^4-5370 q^5-18954 q^6-27760 q^7+\cdots \\
l_{6,a}(\tau)=&q+32 q^2+243 q^3+1024 q^4+3630 q^5+7776 q^6+32936 q^7+\cdots\\
l_{6,b}(\tau)=&q - 32 q^2 - 243 q^3 + 1024 q^4 + 5766 q^5 + 7776 q^6 + 72464 q^7 +\cdots \\
l_{6,c}(\tau)=&q - 32 q^2 + 243 q^3 + 1024 q^4 - 11730 q^5 - 7776 q^6 - 50008 q^7 +\cdots\ .
\end{align}
\end{itemize}
\smallskip

\noindent To summarise, $M_k(\Gamma_0(N))$ is thus generated by
\begin{equation}
\begin{array}{lll}
k=2 : \qquad &  \psi^{(n)}(\tau) \qquad & \hbox{for}\ n|N\,, \ n>1\\
                        & E_2^{\chi_m}(n\tau)  & \hbox{for}\  n m^2|N\,, \ m>1 \\
                        & f_{M,a}(n\tau),\ f_{M,b}(n\tau),\ \ldots \qquad & \hbox{for}\  n M|N\\
                        & & \\
k=4: \qquad   & E_4(n\tau) & \hbox{for}\  n|N            \\
	               & E_4^{\chi_m}(n\tau) &  \hbox{for}\  nm^2|N\,, \ m>1 \\
	               & g_{M,a}(n\tau),\ g_{M,b}(n\tau),\ \ldots \qquad & \hbox{for}\  n M|N\ ,
\end{array}
\end{equation}	
and the cases $k=6,8,\ldots$ {\it etc.} are similar to $k=4$, with $g_{M,a}$ replaced by $h_{M,a}$, {\it etc}.

\section{The modular forms of weight two}\label{a:weightwo}

With the conventions of the previous appendix we can now give the explicit expressions
for the modular forms $F_g\in M_2(\Gamma_0(Nh))$, as well as for $F_g^{h}\in M_{2h}(\Gamma_0(N))$.
For the classes in \eqref{set1} for which $h=1$, the relevant formulae are  \cite{Cheng}
\begin{align}
&F_{{\rm 2A}}(\tau)=16\psi^{(2)}(\tau)\notag\\
&F_{{\rm 3A}}(\tau)=9\psi^{(3)}(\tau)\notag\\
&F_{{\rm 4B}}(\tau)=-4\psi^{(2)}(\tau)+8\psi^{(4)}(\tau)\notag\\
&F_{{\rm 5A}}(\tau)=5\psi^{(5)}(\tau)\\
&F_{{\rm 6A}}(\tau)=-2\psi^{(2)}(\tau)-3\psi^{(3)}(\tau)+6\psi^{(6)}(\tau)\notag\\
&F_{{\rm 7AB}}(\tau)=\frac{7}{2}\psi^{(7)}(\tau)\notag\\
&F_{{\rm 8A}}(\tau)=-2\psi^{(4)}(\tau)+4\psi^{(8)}(\tau)\ ,\notag\\
&F_{{\rm 11A}}(\tau)=\frac{11}{5}\bigl(\psi^{(11)}(\tau) -\eta(\tau)^2\eta(11\tau)^2\bigr)\notag\\
&F_{{\rm 14AB}}(\tau)=\frac{1}{6}\bigl(-2\psi^{(2)}(\tau)- 7\psi^{(7)}(\tau)+14\psi^{(14)}(\tau)
-14\eta(\tau)\eta(2\tau)\eta(7\tau)\eta(14\tau)\bigr)\notag\\[6pt]
&F_{{\rm 15AB}}(\tau)=\frac{1}{8}\bigl(-3\psi^{(3)}(\tau)- 5\psi^{(5)}(\tau)+15\psi^{(15)}(\tau)
-15\eta(\tau)\eta(3\tau)\eta(5\tau)\eta(15\tau)\bigr)\notag\\[6pt]
&F_{{\rm 23AB}}=\frac{23}{22}\bigl(\psi^{(23)}(\tau)- f_{23,a}(\tau)+3f_{23,b}(\tau)\bigr)\ .\notag
\end{align}
In the remaining cases we have found --- the formulae for $F_{\rm 2B}$ and $F_{\rm 4A}$ were
already obtained in \cite{Cheng}
\begin{itemize}
\item $g=$ 2B, $h=2$%, $\Gamma_0(2|2)$
\begin{align}
&F_{{\rm 2B}}(\tau)=-24\psi^{(2)}(\tau)+16\psi^{(4)}(\tau)& \in M_2(\Gamma_0(4))\\
&F_{{\rm 2B}}(\tau)^2=-16E_4(\tau)+256E_4(2\tau)& \in M_4(\Gamma_0(2))
\end{align}
\item $g=$ 3B, $h=3$%, $\Gamma_0(3|3)$
\begin{align}
&F_{3B}(\tau)=-6\psi^{(3)}(\tau)+\frac{9}{2}\psi^{(9)}(\tau)-\frac{9}{2}E_2^{(9)}(\tau)& \in M_2(\Gamma_0(9))\\
&F_{3B}(\tau)^3=-1944\psi^{(3)}(\tau)^3+\frac{45}{2}E_6(\tau)-\frac{2187}{2}E_6(3\tau)&\in M_6(\Gamma_0(3))
\end{align}
%where
%\be E_6(\tau)=-\frac{1}{504}+\sum_{n=1}^\infty \Bigl(\sum_{d|n}d^5\Bigr)q^n\ ,
%\ee
%is the usual Eisenstein series of weight $6$.
\item $g=$ 4A, $h=2$%, $\Gamma_0(4|2)$
\begin{align}
&F_{{\rm 4A}}(\tau)=4\psi^{(2)}(\tau) -12\psi^{(4)}(\tau) +8\psi^{(8)}(\tau)&\in M_2(\Gamma_0(8))\\
&F_{{\rm 4A}}(\tau)^2= -16E_4(2\tau)+ 256E_4(4\tau)&\in M_4(\Gamma_0(4))
\end{align}
\item $g=$ 4C, $h=4$%, $\Gamma_0(4|4)$
\begin{align}
F_{{\rm 4C}}(\tau)&= 2\psi^{(4)}(\tau) -6\psi^{(8)}(\tau) +4\psi^{(16)}(\tau) -4 E^{(16)}_2(\tau)
 &&\in M_2(\Gamma_0(16))\\[6pt]
% F_{4C}(\tau)^2&=16E_4(2\tau) -288E_4(4\tau)+ 512E_4(8\tau) -8 h_8(\tau) &&\in M_4(\Gamma_0(8))\ ,\notag\\[6pt]
F_{{\rm 4C}}(\tau)^4&= -\tfrac{32}{17}E_8(2\tau)+\tfrac{8192}{17}E_8(4\tau)
-16k_2(\tau)-\tfrac{512}{17}k_2(2\tau)  &&\in M_8(\Gamma_0(4))
\end{align}
\item $g=$ 6B, $h=6$%, $\Gamma_0(6|6)$
\begin{align}
&F_{{\rm 6B}}(\tau)=-\tfrac{3}{2}\psi^{(2)}(\tau)-2\psi^{(3)}(\tau)+\psi^{(4)}(\tau)+6\psi^{(6)}(\tau)+
\tfrac{3}{2}\psi^{(9)}(\tau)-4\psi^{(12)}(\tau)\\&
\qquad\qquad\quad -\tfrac{9}{2}\psi^{(18)}(\tau)+3\psi^{(36)}(\tau)
-\tfrac{3}{2}E_2^{(9)}(\tau)\notag\\&
\qquad\qquad\quad  -9E_2^{(9)}(2\tau)-12E_2^{(9)}(4\tau) \hspace{4cm}
 \in M_2(\Gamma_0(36))\notag\\[6pt]
&F_{{\rm 6B}}(\tau)^6=c_1E_{12}(\tau)+c_2E_{12}(2\tau)+c_3E_{12}(3\tau)+c_4E_{12}(6\tau)+c_5\Delta(\tau)
 \\&\qquad\qquad\quad+c_6\Delta(2\tau)+c_7\Delta(3\tau)+c_8\Delta(6\tau)+c_9l_{6,a}(\tau)\notag\\
 &\qquad\qquad\quad+c_{10}l_{6,b}(\tau)+c_{11}l_{6,c}(\tau)
 +c_{12}l_{3}(\tau)+c_{13}l_{3}(2\tau)
 \qquad\in M_{12}(\Gamma_0(6))\notag\end{align}
where $\Delta(\tau)=((240 E_4(\tau))^3-(504E_6(\tau)^2))/1728$ and
%\begin{equation}\begin{array}{|c|c||c|c||c|c|}\hline
% i & c_i & i & c_i & i & c_i\\\hline&&&&&\\[-10pt]
% 1 & \frac{1}{22951565} & 6 &-\frac{91776}{3455} & 11 &-\frac{95}{52}\\[4pt]
% 2 &-\frac{4096}{22951565} & 7 & \frac{37712628}{58735}& 12 &-\frac{194697}{71978}\\[4pt]
% 3 &-\frac{531441}{22951565} & 8 &\frac{27713664}{3455} & 13 &\frac{304128}{2117}\\[4pt]
% 4 &\frac{2176782336}{22951565} & 9 &-\frac{297}{140} &  &\\[4pt]
% 5 &-\frac{189277}{58735} & 10 &-\frac{308}{145} &  &\\[4pt]\hline
%\end{array}
%\end{equation}
\begin{center}
\begin{tabular}{|c|ccccccc|}\hline
 $p$ & 1 & 2 & 3 & 4 & 5 & 6 & 7 \\ \hline &&&&&&&\\[-10pt]
 $c_p$ &
 $\tfrac{1}{22951565}$ & $-\tfrac{4096}{22951565}$ & $-\tfrac{531441}{22951565}$ & $\tfrac{2176782336}{22951565}$ & $-\tfrac{189277}{58735}$ & $-\tfrac{91776}{3455}$ & $\tfrac{37712628}{58735}$ \\[6pt]
 \hline\hline $p$ & 8 & 9 & 10 & 11 & 12 & 13&\\
 \hline &&&&&&&\\[-10pt]
 $c_p$ &
 $\tfrac{27713664}{3455}$ & $-\tfrac{297}{140}$ & $-\tfrac{308}{145}$ & $-\tfrac{95}{52}$
 & $-\tfrac{194697}{71978}$ & $\tfrac{304128}{2117}$&\\[6pt] \hline
\end{tabular}
\end{center}
\vspace{0.2cm}
\item $g=$10A, $h=2$%, $\Gamma_0(10|2)$
\begin{align}
F_{{\rm 10A}}(\tau)&=\psi^{(2)}(\tau)-\frac{2}{3}\psi^{(4)}(\tau)+\frac{5}{3}\psi^{(5)}(\tau)-5\psi^{(10)}(\tau)
\\&\qquad+\frac{10}{3}\psi^{(20)}(\tau)\notag
-\frac{10}{3}\eta(2\tau)^2\eta(10\tau)^2 &\in M_2(\Gamma_0(20))\\
F_{{\rm 10A}}(\tau)^2&=\frac{1}{39}E_4(\tau)-\frac{16}{39}E_4(2\tau)-\frac{625}{39}E_4(5\tau)+\frac{10000}{39}E_4(10\tau)
\\&\qquad-\frac{10}{3}g_{10}(\tau)-\frac{35}{13}g_{5}(\tau)+\frac{40}{13}g_{5}(2\tau) &\in M_4(\Gamma_0(10))\notag
\end{align}
\item $g=$ 12A, $h=2$%, $\Gamma_0(12|2)$
\begin{align}
F_{{\rm 12A}}(\tau)&=-\frac{1}{2}\psi^{(2)}(\tau)+\frac{3}{2}\psi^{(4)}(\tau)+\frac{3}{2}\psi^{(6)}(\tau)-\psi^{(8)}(\tau)
\\&\qquad-\frac{9}{2}\psi^{(12)}(\tau)+3\psi^{(24)}(\tau)-3f_{24}(\tau) &\in M_2(\Gamma_0(24))\notag\\
F_{{\rm 12A}}(\tau)^2&=\frac{1}{5}E_4(2\tau)-\frac{16}{5}E_4(4\tau)-\frac{81}{5}E_4(6\tau)+\frac{1296}{5}E_4(12\tau)
\\&\qquad-3g_{12}(\tau)-3g_6(\tau)+\frac{24}{5}g_6(2\tau) &\in M_4(\Gamma_0(12))\notag
\end{align}
\item $g=$ 12B, $h=12$%, $\Gamma_0(12|2)$
\begin{align}
F_{{\rm 12B}}(\tau)&=\frac{1}{8}\psi^{(4)}(\tau)-\frac{3}{8}\psi^{(8)}(\tau)-\frac{1}{2}\psi^{(12)}(\tau)
+\frac{1}{4}\psi^{(16)}(\tau)+\frac{3}{2}\psi^{(24)}(\tau)+\frac{3}{8}\psi^{(36)}(\tau)\notag
\\&-\psi^{(48)}(\tau)
-\frac{9}{8}\psi^{(72)}(\tau)+\frac{3}{4}\psi^{(144)}(\tau)-\frac{3}{2}E_2^{(9)}(4\tau)-9E_2^{(9)}(8\tau)
-12E_2^{(9)}(16\tau)\notag\\&-\frac{1}{4}E_2^{(16)}(\tau)-3E_2^{(16)}(3\tau)-\frac{27}{4}E_2^{(16)}(9\tau)
-\frac{3}{4}E_2^{(144)}(\tau)-\frac{3}{2}f_{24}(2\tau)\notag\\&-\frac{27}{2}f_{24}(6\tau)-\frac{3}{4}f_{48}(\tau)
+\frac{27}{4}f_{48}(3\tau)+\frac{9}{2}f_{72}(2\tau)-\frac{9}{4}f_{144,b}(\tau) \ \in M_2(\Gamma_0(144))\notag\\
\end{align}
We have also worked out the expansion of $F_{{\rm 12B}}^{12}$ in terms of generators of $M_{24}(\Gamma_0(12))$, but since the final expression is exceedingly complicated, we shall not spell it out here.
\item $g=$ 21AB, $h= 3$%, $\Gamma_0(21|3)$\\
\begin{align} F_{{\rm 21AB}}(\tau)=
&\frac{1}{8}\psi^{(3)}(\tau)+\frac{7}{32}\psi^{(7)}(\tau)-\frac{3}{32}\psi^{(9)}(\tau)-\frac{7}{8}\psi^{(21)}(\tau)+
\frac{21}{32}\psi^{(63)}(\tau)+\frac{3}{32}E_2^{(9)}(\tau)\notag\\
&-\frac{147}{32}E_2^{(9)}(7\tau)-\frac{21}{32}f_{21}(\tau)
+\frac{189}{32}f_{21}(3\tau)-\frac{63}{32}f_{63,a}(\tau)
%\frac{1}{32}\psi^{(3)}(\tau)
%-\frac{49}{8}\psi^{(3)}(7\tau)-\frac{9}{32}\psi^{(7)}(3\tau)+\frac{27}{32}\psi^{(7)}(9\tau)
%-\frac{9}{2}E_2^{(9)}(\tau)\\
%&\qquad-\frac{21}{32}f_{21}(\tau)+\frac{189}{32}f_{21}(3\tau)-\frac{63}{32}f_{63}(\tau)+\frac{189}{16}\psi^{(3)}(21\tau)
%+\frac{147}{32}\psi^{(3)}(7\tau)
\qquad\in M_2(\Gamma_0(63))\\[10pt]
F_{{\rm 21AB}}(\tau)^3=&%-45E_4(\tau)\psi^{(3)}(\tau)-105E_4(\tau)\psi^{(7)}(\tau)+315 E_4(\tau)\psi^{(21)}(\tau)
%-\tfrac{63}{4}E_6(\tau)
-\tfrac{1}{169936}\,E_6(\tau)+\tfrac{729}{169936}\,E_6(3\tau)
+\tfrac{117649}{169936}\,E_6(7\tau)-\tfrac{85766121}{169936}E_6\,(21\tau)\notag\\[6pt]
& -\tfrac{85641}{36608}\,h_{3}(\tau)-\tfrac{352947}{36608}\,h_{3}(7\tau)-\tfrac{64491}{94600}\,h_{7a}(\tau)
-\tfrac{11263}{9728}\,h_{7b}(\tau)-\tfrac{18277}{4864}\,h_{7c}(\tau)\notag\\[6pt]
&-\tfrac{2740311}{189200}\,h_{7a}(3\tau)+\tfrac{66339}{9728}\,h_{7b}(3\tau)
-\tfrac{250047}{4864}\,h_{7c}(3\tau)
-\tfrac{441}{352}\,h_{21a}(\tau)\notag\\[6pt]
&-\tfrac{1323}{8800}\,h_{21b}(\tau)-\tfrac{735}{512}\,h_{21c}(\tau)-\tfrac{3087}{6400}\,h_{21d}(\tau)
%
%
%&\qquad+\frac{26019}{352}g_{21,a}(\tau)-\frac{953883}{8800}g_{21,b}(\tau)
%-\frac{18375}{512}g_{21,c}(\tau)-\frac{373527}{6400}g_{21,d}(\tau)\notag\\&\qquad-\frac{55412679}{2816}g_{3}(7\tau)
%+\frac{1331883}{4400}g_{7,a}(3\tau)-\frac{3230199}{512}g_{7,b}(3\tau)-\frac{1893213}{256}g_{7,c}(3\tau)\notag\\ &\qquad
%-\frac{44997}{2816}g_{3}(\tau)+\frac{13923}{2200}g_{7,a}(\tau)-\frac{44317}{512}g_{7,b}(\tau)-\frac{147343}{256}g_{7,c}(\tau)
\qquad\qquad\in M_6(\Gamma_0(21))\ .\end{align}
\end{itemize}


\begin{thebibliography}{99}
\bibitem{EOT} T.~Eguchi, H.~Ooguri and Y.~Tachikawa,
{\it Notes on the K3 surface and the Mathieu group $M_{24}$},
Exper.\ Math.\  {\bf 20}, 91 (2011)
{\tt [arXiv:1004.0956 [hep-th]]}.
  %%CITATION = ARXIV:1004.0956;%

\bibitem{EH}
T.~Eguchi  and K.~Hikami,
{\it Superconformal algebras and mock theta functions 2. Rademacher
expansion for K3 surface},
Commun.\ Number Theory Phys. {\bf 3}, 531 (2009)
{\tt [arXiv:0904.0911 [math-ph]]}.

\bibitem{FLM}
I.B.~Frenkel, J.~Lepowsky and A.~Meurman,
{\it Vertex Operator Algebras and the Monster},
Pure and Applied Math.\ {\bf 134}, Academic Press  (1988).

\bibitem{Thompson}
J.G.~Thompson,
{\it Some numerology between the Fischer-Griess Monster and the elliptic
modular function},
Bull.\ Lond.\ Math.\ Soc.\  {\bf 11}, 352 (1979).

\bibitem{CN} J.H.~Conway and S.~Norton, {\it Monstrous moonshine},
Bull.\ Lond.\ Math.\ Soc.\ {\bf 11}, 308 (1979).

\bibitem{Gannon} T.~Gannon, {\it Moonshine beyond the Monster:
The Bridge connecting Algebra, Modular Forms and Physics},
Cambridge University Press (2006).

\bibitem{Cheng}
M.C.N.~Cheng,
{\it K3 Surfaces, N=4 dyons, and the Mathieu group $M_{24}$,}
Commun.\ Number\ Theory\ Phys. {\bf 4}, 623 (2010)
{\tt [arXiv:1005.5415 [hep-th]]}.
%%CITATION = ARXIV:1005.5415;%%

\bibitem{Gaberdiel:2010ch}
M.~R.~Gaberdiel, S.~Hohenegger, R.~Volpato,
  {\it Mathieu twining characters for K3,}
  JHEP {\bf 1009}, 058 (2010)
  {\tt [arXiv:1006.0221 [hep-th]].}
%%CITATION = ARXIV:1006.0221;%%

\bibitem{Kawai:1993jk}
T.~Kawai, Y.~Yamada, and S.-K. Yang,
{\it Elliptic genera and $\N=2$ superconformal field theory},
Nucl.\ Phys.\  B {\bf 414}, 191 (1994) {\tt [arXiv:hep-th/9306096]}.

\bibitem{David:2006ji}
J.R.~David, D.P.~Jatkar and A.~Sen,
{\it Product representation of dyon partition function in CHL models},
JHEP {\bf 0606}, 064 (2006)
{\tt [arXiv:hep-th/0602254]}.
 %%CITATION = JHEPA,0606,064;%%

\bibitem{EichlerZagier}
M.~Eichler and D.~Zagier,
{\it The Theory of Jacobi Forms},
Birkh\"auser (1985).

\bibitem{EOTY}
T.~Eguchi, H.~Ooguri, A.~Taormina and S.K.~Yang,
{\it Superconformal algebras and string compactification on manifolds
with $SU(N)$ holonomy},
Nucl.\ Phys.\  B {\bf 315}, 193 (1989).
%%CITATION = NUPHA,B315,193;%%


\bibitem{Eguchi:1987wf}
T.~Eguchi and A.~Taormina,
{\it Character formulas for the $\N=4$ superconformal algebra},
Phys.\ Lett.\  B {\bf 200}, 315 (1988).
%%CITATION = PHLTA,B200,315;%%

\bibitem{Eguchi:1988af}
T.~Eguchi and A.~Taormina,
{\it On the unitary representations of $N=2$ and $N=4$ superconformal algebras},
Phys.\ Lett.\ B {\bf 210}, 125 (1988).
%%CITATION = PHLTA,B210,125;%%

\bibitem{Mukai}
S.~Mukai,
{\it Finite groups of automorphisms of $K3$ surfaces and the Mathieu group},
Invent.\ Math. {\bf 94}, 183 (1988).

\bibitem{Kondo}
S.~Kondo,
{\it Niemeier lattices, Mathieu groups and finite groups of symplectic automorphisms
of K3 surfaces},
Duke Math.\ Journal {\bf 92}, 593 (1998), appendix by S.~Mukai.

\bibitem{FMN}
D.~Ford, J.~McKay and S.~Norton,
{\it More on replicable functions},
Commun.\ Algebra {\bf 22}, 5175 (1994).


\bibitem{Taormina:2010pf}
A.~Taormina and K.~Wendland,
{\it The symmetries of the tetrahedral Kummer surface in the Mathieu group $\MM_{24}$},
{\tt arXiv:1008.0954 [hep-th]}.
  %%CITATION = ARXIV:1008.0954;%%

\bibitem{pro}
M.R.~Gaberdiel, S.~Hohenegger and R.~Volpato,
work in progress.

\bibitem{webInteger} {\tt http://www.research.att.com/$\tilde{\phantom{w}}$njas/sequences/}

\bibitem{Lang}
S.~Lang, {\it Introduction to Modular Forms},
Grundlehren der Mathematischen Wissenschaften
{\bf 222}, Springer Verlag, Berlin (1995).

\bibitem{Stein}
W.~Stein, {\it Modular Forms, a Computational Approach},
Graduate Studies in Mathematics {\bf 79},
{American Mathematical Society},
Providence, RI  (2007).

\bibitem{Miyake}
T.~Miyake, {\it Modular Forms}, Springer Verlag,
Berlin (2006).

\bibitem{EHT}
  T.~Eguchi, K.~Hikami,
  {\it Note on Twisted Elliptic Genus of K3 Surface,}
  Phys.\ Lett.\  B {\bf 694}, 446 (2011)
  {\tt [arXiv:1008.4924 [hep-th]].}
\end{thebibliography}
\end{document}